# Theoretical Study of $C_{36}$ and $O@C_{36}$ Fullerene Isomers with ($C_1$, $C_2$, $C_s$, $D_2$, $D_{3h}$) symmetries: Geometry Optimization, Electrical Properties and Spectroscopic Analysis

Fouad Aljneed[1], Nabil Joudieh[1], Issam Aljghami[1], Khansaa Hussein[2] and Nidal Chamoun[3,4,*]

[1]Physics Department, Faculty of sciences, Damascus University, P.O. Box 30621, Damascus, Syria
[2]Chemistry Department, Faculty of sciences, Damascus University, P.O. Box 30621, Damascus, Syria
[3] Statistics Department, Faculty of sciences, Damascus University, P.O. Box 30621, Damascus, Syria
[4]CASP, Antioch Syrian Private University, Maaret Saidnaya, Damascus, Syria
Emails: fouad.aljneed@damascusuniversity.edu.sy, nabil.joudieh@damascusuniversity.edu.sy, issam.aljghami@damascusuniversity.edu.sy, khansaa1.hussein@damascusuniversity.edu.sy,

*corresponding author: nidal_chamoun@yahoo.com, chamoun@uni-bonn.de

**Abstract.** In this study, we conducted a theoretical analysis of specific $C_{36}$ and $O@C_{36}$ Fullerene isomers, namely those with ($D_{3h}$, $C_1$, Cs, $C_2$, $D_2$) symmetries, in the gaseous phase using the DFT method at B3LYP/6-31G* level. We studied geometry optimizations, relative stability, atomization energies, Fermi energy, energy gap, electronic properties, electric dipole moment, polarizabilities, and thermodynamic analysis, along with IR and NMR spectra. Consistently, our findings revealed distinct properties and characteristics among different $C_{36}$ fullerene isomers. The energetic order of $C_{36}$ fullerene isomers was established as $C_{36}$-$D_{3h}$ < $C_{36}$-$C_1$ < $C_{36}$-Cs < $C_{36}$-$C_2$ < $C_{36}$-$D_2$, a trend unaffected by the encapsulation of oxygen. Notably, the $D_2$ isomer displayed the smallest energy gap, indicating higher electrical conductivity compared to other isomers, while it exhibited the largest gap after encapsulation. Furthermore, we observed that the $C_1$ ($D_2$) isomer exhibited the largest (smallest) Electric dipole moment among the studied $C_{36}$ isomers, whereas the $C_2$ ($D_2$) isomer demonstrated the largest Electric dipole moment among the $O@C_{36}$ isomers studied. The encapsulation of oxygen in $C_{36}$ isomers influenced their properties, including alterations in electronic properties, IR, and NMR chemical shifts. A detailed analysis of BSSE corrections showed the tininess of their impact on the analyzed observables, such that uncorrected and BSSE-corrected calculations led to nearly identical results, confirming the robustness and reliability of the B3LYP/6-31G* approach used in this work and, eventually, in other large and rigid fullerenes systems.

## 1. Introduction

The discovery of fullerenes has spurred active research in the realm of endohedral fullerenes (EFs), involving the encapsulation of atoms, ions, molecules and small clusters [1]. EFs have garnered attention from physicists and chemists owing to the profound impact of the encapsulated entity on the geometric, electronic and optical properties of fullerenes [2]. This has paved the way for compounds with potential applications in quantum computing, magnetic materials, biomedicine and pharmacology [3]. However, experimental studies of smaller fullerenes pose challenges due to their reactivity and high curvature, primarily stemming from adjacent pentagonal rings [2, 4-5]. Among small fullerenes, $C_{36}$ stands out as the only one isolated in the solid phase, boasting 15 conventional isomers. It emerged as one of the first "magic-number" small fullerenes identified via mass spectroscopy [6-7]. Notably, extensive research has been dedicated to $C_{36}$, including milestone achievements such as the stable synthesis of a lower $C_{36}$ fullerene by Piskoti and Zettl [8] and the generation of $C_{36}$ clusters through appropriate non-equilibrium growth conditions by Piskoti et al. [6]. Grossman et al. [9] conducted a comprehensive analysis of the electronic and structural properties of various $C_{36}$ fullerene isomers, using a pseudopotential DFT approach alongside other calculations. In a separate study, Cote et al. [10] delved into the electron-phonon interaction in solid $C_{36}$. Furthermore, Halac et al. [11] explored the molecular structures and vibrational properties of newly synthesized $C_{36}$ fullerene using semi-empirical potential calculations. Fowler et al. [12] conducted theoretical investigations into the structure and energies of $C_{36}$-based fullerenes, hydrides, oligomers and solids at the DFTB level. Their work accounted for the phenomenon where $C_{36}$ forms stronger inter-cage bonds compared to larger fullerenes. In a separate study, Jishi and Dresselhaus [13] utilized first-principle calculations to examine the vibrational infrared and Raman-active modes of neutral, positively and negatively charged $C_{36}$ molecules. They also explored the electronic structure, structural deformation, charge distributions, electron densities and spin densities, using the B3LYP method for anionic and di-anionic species of the promising $C_{36}$ fullerene configurations $D_{2d}$ and $C_{2v}$.

These studies collectively provide valuable insights into the molecular properties of $C_{36}$ fullerene. In 2001, Beu et al. [15] utilized tight-binding molecular dynamics to investigate the structural and vibrational properties of $C_{36}$ and its oligomers. Varganov et al. [16] focused on studying the five lowest energy isomers of $C_{36}$ using the CASSCF method alongside single and multireference MP2 theories to elucidate the effect of electron correlation on energies, HOMO and LUMO energies. Furthermore, the binding energies, relative stabilities and basis set effects of $D_{6h}$, $D_{3h}$ and $D_{2d}$ isomers of $C_{36}$ were explored using Hartree-Fock, DFT, MP2 and coupled cluster methods [17]. Jin and Hao [18] presented the Stone-Wales transformation in $C_{36}$, while the relative energies and HOMO-LUMO energies of all $C_{36}$ isomers were analyzed with respect to basis sets in another study [19]. Additionally, the SCC-DFTB method was employed to compute the geometrical, electronic and vibrational properties of $C_{36}$ isomers [20]. In the realm of $C_{36}$ fullerene research, various investigations have shed light on its interactions and properties. A study by [5] explored the interaction between $C_{36}$ fullerene and the glycine radical using the DFT method. In 2018, Yong Ma et al. [21] reported on distinguishing the six stable $C_{36}$ fullerene isomers through a combined approach involving DFT and X-ray photoelectron spectroscopy. In 2024, Idrissi, S et al. reported an investigation of the magnetic properties of the $C_{36}$ fullerene structure using Monte Carlo simulations [22]. Concurrently, studies on endohedral fullerene X@$C_{36}$ were developed alongside those focusing on $C_{36}$ isomers. In the gas phase, numerous X@$C_{36}$ species (X = Th, Sc, Pr, Nd, Tb, Dy, Ho, Er, Lu [22], U [2], La, Gd, Y, Ce [23,24]) were experimentally discovered. The earliest theoretical investigation into the electronic and structural properties of $C_{36}$ and its endohedral compounds dates back to 1998 by Grossman et al. [9]. Employing pseudopotential DFT, this study examined several $C_{36}$ isomers and their endohedral counterparts. Notably, they calculated the endohedral binding energies of various X@$C_{36}$ molecules and studied the chemical shifts of the two lowest-energy fullerene isomers, $D_{6h}$ and $D_{2d}$. Their findings suggested that $C_{36}$ is likely the smallest fullerene capable of encapsulating a wide range of atoms. Subsequently, other $C_{36}$ endohedral fullerenes have been theoretically studied, encompassing a diverse array of encapsulated atoms, including H, K, Li, Na [25], Be, Mg, Ca [26], He, N [27], Mo [28], C, F [29,30], Ti, V, Cr, Mn, Fe, Co, Ni, Cu [31], U [2], He, Ne, Ar [32,33], La, Y and Sc [34]. Some of these fullerenes have undergone both theoretical and experimental scrutiny, exemplified by U@$C_{36}$ [2], Y@$C_{36}$, Sc@$C_{36}$, and La@$C_{36}$ [35].

In this work, we investigated the stability and properties of five isomers of $C_{36}$ fullerene, namely those with $C_1$, $C_2$, Cs, $D_2$ and $D_{3h}$ symmetries with and without encapsulating an oxygen atom to form O@$C_{36}$. Although there are more stable isomers with different symmetries, well examined in the literature [16, 17, 36, 37], however we opted to concentrate our work on less studied ones analyzing first, to our knowledge, the possibilities of oxygen insertion. The choice of oxygen as the atomic endohedral dopant stemmed from its high electronegativity, its ability to undergo oxidation with five oxidation states and its potential application as a directed therapeutic agent in medical contexts, such as drug carriers. Additionally, the study of endohedral N@$C_{60}$ as a fundamental unit for quantum computing [35] underscores the potential utility of O@$C_{36}$ in this domain. The amalgamation of an encapsulated atom with fullerene in endohedral X@$C_{36}$ structures presents an intriguing phenomenon, yielding unique properties stemming from their interaction. Within this framework, the fullerene cage acts as a protective Faraday cage, preserving the spin states of the encapsulated atom while exhibiting weak interaction with it. Additionally, the smaller size of the oxygen atom, relative to nitrogen, presents an advantage for potential applications.

In the present study, we utilized the DFT/B3LYP method in conjunction with the 6-31G* basis set implemented in the ORCA 5.0.1 program [38] in order to explore the stability and properties of the five $C_{36}$ fullerene isomers and their endohedral counterparts, O@$C_{36}$. Through this approach, we scrutinized various aspects including optimized geometry and relative stability, energy and thermodynamic quantities, theoretical IR/NMR spectra and polarizability of the compounds.

Apart from the relative, gap and Fermi energies in the $C_{36}$ fullerenes, which were estimated in [19, 20], no theoretical/experimental data for the other observables in the studied isomers, and -a fortiori- in their -not yet fabricated- endohedral counterparts, is available in the literature in order to assess the uncertainty of our results. Thus, our tables, except Table 2 (3) related to relative (gap and Fermi) energy, do not include comparison with other data.

A critical aspect of computational quantum chemistry is the potential for Basis Set Superposition Error (BSSE), which can artificially stabilize molecular systems when moderate basis sets are used. In order to address this point, we performed comparative calculations on one isomer and its endohedral derivative, namely $C_{36}$-$D_{3h}$ and O@$C_{36}$-$D_{3h}$, using both standard and BSSE-corrected (counterpoise) methods. Our results show that BSSE corrections have a negligible impact on the computed geometries, energies, and electronic properties, while differences in bond lengths, angles, and atomic charges were minimal -typically less than 0.4%-, whereas the electron density distribution remained virtually unchanged. This confirms that the B3LYP/6-31G* approach is robust and reliable for modeling large, rigid fullerenes and their endohedral derivatives, and that explicit BSSE correction is unnecessary in such contexts.

As a result, this work not only advances the understanding of the structural and electronic behavior of $C_{36}$ and $O@C_{36}$ isomers but also validates the computational methodology for future studies of large, rigid fullerenes, where computational efficiency and accuracy are both essential.

The plan of the paper is as follows. In section 2 we state the computational tools used in the study and argue for their validity, whereas we present the results in section 3 which is divided into several subsections. In subsection 3.1 we study the structure and energy levels of the compounds, while in subsection 3.2 (3.3) we study the electronic (thermodynamic) properties. Vibration and IR Spectra are presented in subsection 3.4, whereas the NMR one is presented in 3.5. We conclude with a summary in section 4. An appendix with three sections includes details of the BSSE analysis comparing calculations with and without it, and of the convergence tests in ORCA.

## 2. Computational methodology

### 2.1. Method & limitations

The calculations, including geometry optimizations and others, were carried out using the energy gradient that was constructed analytically and implemented in the ORCA 5.0.1 program package [38]. DFT/B3LYP method with a standard 6-31G* basis set was used for the calculations.

Although one is aware of the limitations of the DFT/B3LYP with a 6-31G* basis set, particularly in relation to the electronic properties of the systems under study, however -as we shall argue- the results obtained are reasonably plausible, and we believe they provide a qualitative and semi-quantitative view of the studied fullerenes.

Actually, B3LYP is a widely used exchange-correlation functional in DFT that employs the hybrid generalized gradient approximation (GGA) method. This functional combines 20% exact Hartree-Fock exchange with 80% DFT exchange-correlation, utilizing the Becke88 (B88) functional for exchange and the Lee-Yang-Parr (LYP) functional for correlation. Its classification as a global hybrid reflects the existence of a non-parametrized portion of exact Hartree-Fock exchange integrated into the GGA framework, enhancing accuracy for properties like reaction barriers and electronic structures compared to pure GGA functionals. B3LYP occupies the fourth rank of Jacob's Ladder of DFT functionals, a categorization reserved for hybrid methods blending exact exchange with GGA or meta-GGA components. This hybrid GGA architecture balances computational efficiency with improved predictive power, contributing to its dominance in computational chemistry for studying molecular geometries, vibrational frequencies, and thermochemical data. While modern functionals have emerged, B3LYP remains a benchmark due to its historical reliability and extensive validation across diverse systems.

Moreover, one can adapt the DFT method easily in ORCA environments to account for BSSE corrections [39], resulting from overlap of neighboring molecular basis functions, by implementing counterpoise procedure [40, 41].

In fact, BSSE is a persistent challenge in computational quantum chemistry stemming from the use of incomplete or moderate, atom-centered basis sets. BSSE artificially stabilizes molecular systems by enabling fragments to "borrow" basis functions, leading to overestimated binding energies. The Counterpoise (CP) method and Chemical Hamiltonian Approach (CHA) are two common strategies employed to mitigate BSSE.

**2.2. Assessment of the Method**

We argue that one can use the B3LYP/6-31G* as an effective and trustworthy choice for studying large and complex fullerene systems, with minimal concern for BSSE artifacts.

As a matter of fact, and to assess the suitability of the selected basis set and the B3LYP/6-31G* approach for studying fullerenes and their endohedral derivatives, as well as to determine the need for BSSE corrections, we conducted calculations on $C_{36}$-$D_{3h}$ and O@$C_{36}$-$D_{3h}$ using both uncorrected and BSSE-corrected methods employing B3LYP/6-31G* and B3LYP-gCP/6-31G* techniques. A summary of BSSE corrections and their relevance to this study is provided in the appendix, where we examined the influence of BSSE corrections on the geometric/energetic characteristics and other observables. The computational analysis shows that the relative changes in bond lengths (e.g., $C_{25}$-$O_{37}$; $C_{22}$-$O_{37}$) and valence angles (e.g., $C_{14}$-$O_{37}$-$C_{15}$; $C_{25}$-$O_{37}$-$C_{22}$) upon performing the BSSE corrections are negligible (0.31%, 0.37%, 0.21%, 0.19% respectively). These small variations highlight the structural stability of both molecules, indicating that BSSE corrections do not significantly alter equilibrium geometries or distort the potential energy surface. Similarly, energy-related properties remain largely unaffected by BSSE, implying that the artificial stabilization caused by basis set overlap is minimal in these systems. This stability conforms with the rigid or near-rigid nature of $C_{36}$-$D_{3h}$ and O@$C_{36}$-$D_{3h}$, which is typical of systems with weak non-covalent interactions. The agreement between geometric and energetic results confirms the reliability of the B3LYP/6-31G* method for accurately modeling these structures. Consequently, BSSE corrections may not be necessary when investigating large fullerenes and their endohedral derivatives, especially when computational efficiency is a priority.

Likewise, Mulliken atomic charges for $C_{36}$-$D_{3h}$ and O@$C_{36}$-$D_{3h}$, calculated using both B3LYP/6-31G* and the BSSE-corrected method, are presented in the appendix. The data show that the average atomic charge density modulus for $C_{36}$-$D_{3h}$ decreases from 0.02128 (B3LYP/6-31G*) to 0.018383 (with BSSE correction), while for O@$C_{36}$-$D_{3h}$, it decreased from 0.03113 to 0.030065. Notably, the charge on the oxygen atom changes very little (from 0.4818 to 0.4804 amounting to a 0.29% relative change) after applying BSSE correction. The nearly identical atomic charges for carbon and oxygen atoms, regardless of BSSE correction, demonstrate that BSSE has a minimal effect on electron density distribution. This suggests that artificial stabilization due to basis set overlap does not significantly distort the calculated electron density or atomic charge distribution in these molecules. Therefore, for this rigid system and chosen basis set, the computed atomic charges are robust and reliable concerning BSSE corrections, and the decision to include or omit counterpoise correction is unlikely to influence substantially charge-dependent properties or their interpretations. The limited flexibility and size of the 6-31G* basis set, combined with the structural rigidity of the fullerene cage, once again effectively minimize BSSE effects on electron density-related properties.

Thus, our work detailed in the appendix, supports the suitability of the selected methodology for predicting structural and electronic properties in similar rigid frameworks, in that it shows the B3LYP/6-31G* basis set was sufficiently large for the BSSE corrections to be small, which justified us adopting this not so-powerful combination with no much computing resources.

We stress again that the use of B3LYP/6-31G* strikes an optimal balance between computational cost and accuracy, making it well-suited -even without explicit BSSE corrections- for geometry optimizations of fullerenes, whose quasi-rigid nature means the corresponding geometries and energies are inherently stable and largely resistant to BSSE-induced distortions. As a result, weak non-covalent interactions do not significantly affect the calculated structures or energetics, so that principal electronic characteristics, such as the energies of the frontier orbitals and the dipole moments, are largely unaffected, and the form of the potential energy surface around equilibrium geometries is preserved despite the limitations of the selected basis set.

These findings -on the B3LYP/6-31G* providing a reliable and practical approach for minimizing BSSE effects on both structural and electronic properties- support the broader application of this computational protocol to larger fullerenes and their endohedral derivatives, where the use of more extensive basis sets would be computationally prohibited. Moreover, as fullerenes increase in size, non-covalent interactions become even weaker due to greater distances between non-bonded atoms, further enhancing the reliability of the B3LYP/6-31G* methodology.

### 2.3. Convergence test in the ORCA program

Convergence tests for $C_{36}$-$D_{3h}$ and O@$C_{36}$-$D_{3h}$ fullerenes were systematically performed using the ORCA5 quantum chemistry package, employing both SCF and geometry optimization criteria to ensure reliable electronic and structural results. For these structurally rigid systems, standard convergence tolerances were sufficient, and stricter criteria were unnecessary. Geometry optimizations utilized the cornerstone optimization algorithm, the BFGS quasi-Newton method, within redundant internal coordinates. Two sets of convergence parameters were applied: (1) SCF criteria involving tight energy and density matrix thresholds, and (2) geometry criteria specifying strict energy, gradient, and displacement limits, with Hessian initialization via the Almlöf model. Comparative tests using B3LYP/6-31G* (uncorrected) and B3LYP-gCP/6-31G* (BSSE-corrected) methods revealed that, while the uncorrected approach satisfied all convergence requirements, BSSE corrections led to a failure in the MAX gradient test for both $C_{36}$-$D_{3h}$ and O@$C_{36}$-$D_{3h}$, without altering the tolerance settings. These results confirm that the chosen convergence protocols in ORCA5 provide a robust balance of computational efficiency and accuracy for rigid fullerene systems, and that BSSE corrections have minimal impact on the convergence behavior and consequently on the reliability of structural and electronic predictions.

## 3. Results and Discussion

The objective of our study is to compute and assess diverse properties of various isomers of $C_{36}$ fullerene and their endohedral O@$C_{36}$ compounds. This entails determining optimized electronic energy and geometry, dipole moment, polarizability, along with several energy and thermodynamic parameters. Furthermore, the investigation covers also the examination of vibration frequencies, theoretical infrared spectra and NMR spectra for these compounds.

### 3. 1. Structural properties, total and relative energies

Figure 1 displays the optimized geometries of both the $C_{36}$ fullerene isomers and their endohedral counterpart O@$C_{36}$. These structures adhere to Euler's theorem, which dictates that a fullerene configuration must comprise 8 hexagons and 12 pentagons. The diversity among isomers arises solely from the arrangement of hexagons and pentagons. Notably, the most stable structure is characterized by having the minimal number of adjacent pentagons, since an increasing numbers of these pentagons leads to increasing internal stress.

The findings indicated that the oxygen atom did not occupy the central position within the fullerene cage; rather, it was positioned proximate to two carbon atoms. This positioning led to an extension in the bond length and an increase of the angle between these carbon atoms and their neighboring counterparts, as presented in Table 1. Such asymmetry is attributed to the electronegativity of the oxygen atom and the heightened charge densities of the adjacent carbon atoms.

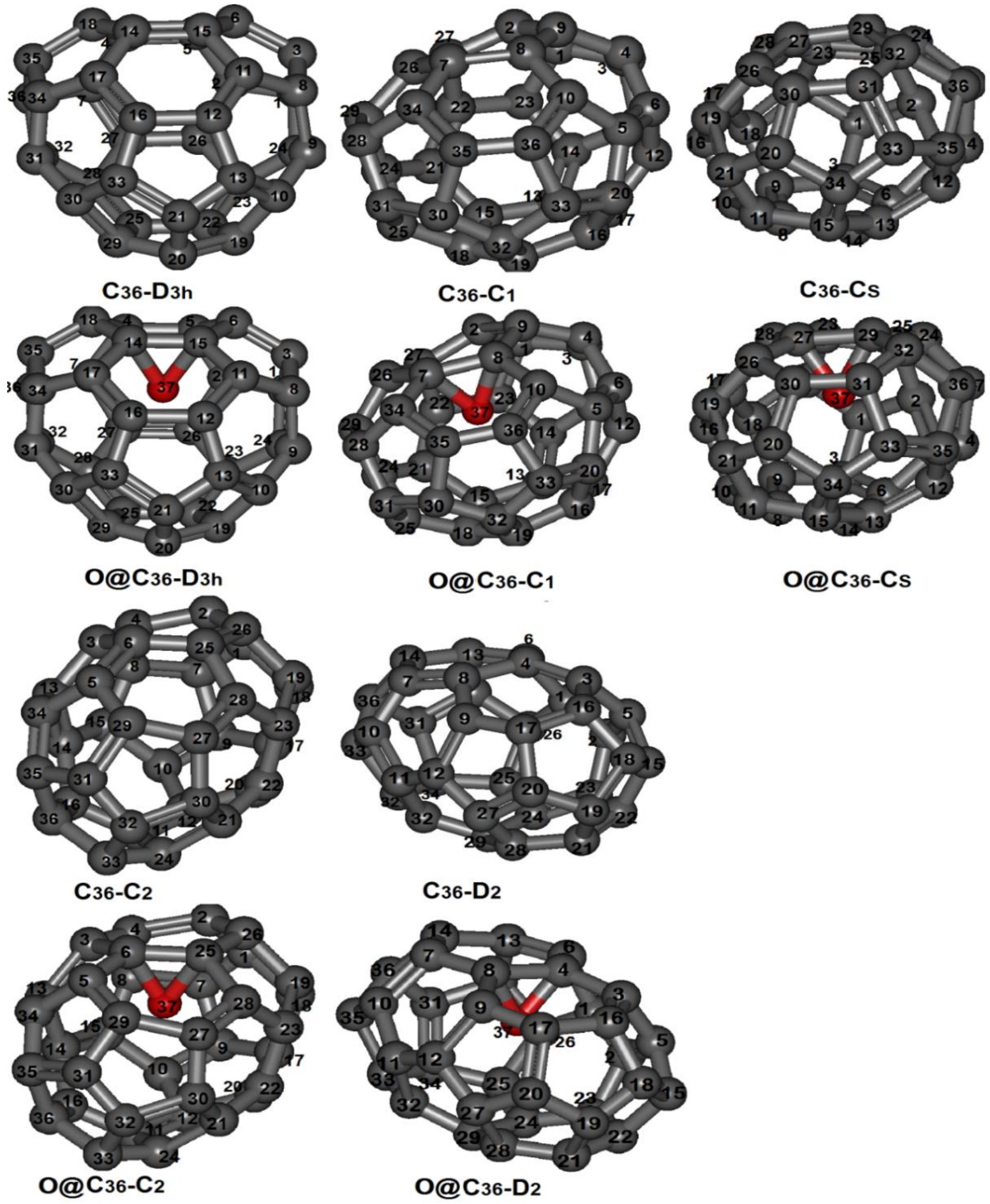

Fig. 1. Optimized structure of $C_{36}$ and O@$C_{36}$ fullerene isomers using the DFT method at B3LYP/6-31G* level. Red color denotes an Oxygene atom[1].

Actually, for the O@$C_{36}$-$D_{3h}$ isomer, calculations, with or without BSSE corrections, in the appendix show that the O atom (with charge around -0.48 e) is at a separating distance of approximately 1.5 A$^o$ from two C atoms denoted 14, 15 (each with charge around + 0.12 e), whereas the same distance to their opposite face atoms denoted 22, 25 (with charge around +0.02 e) was around 3.3 A$^o$, in line with more attractive force between the O-atom and the former C-atoms pair than the latter one. The small size of the $C_{36}$ cluster, exhibited by the separating distance of order 4.48 A$^o$ between the atoms (14 & 25) or between the atoms (15 & 22), may contribute to O being not at center. Moreover, the C atoms pair (14, 15) are positioned within hexagons, in contrast to the other pair (22, 25) positioned within hexagons and pentagons, which may interpret the tendency of the encapsulated O atom to be connected with the former pair, since it is hexagons, rather than pentagons, which help in further distributing the stress leading thus to more stability.

[1] Plots were obtained using the Gabedit software (https://gabedit.sourceforge.net).

Table 1. The bond length (Å), bond angles (°), of $C_{36}$ and O@$C_{36}$ fullerene isomers at their equilibrium geometries using the DFT method at B3LYP/6-31G* level. See Figure 1 for labelling of the atoms.

| Symmetry | Bond length (Å) | | | Bond angle (°) | | |
|---|---|---|---|---|---|---|
| | Definition | $C_{36}$ | O@$C_{36}$ | Definition | $C_{36}$ | O@$C_{36}$ |
| $D_{3h}$ | $C_{14}C_{15}$ | 1.401 | 1.448 | $C_{15}C_{14}C_{17}$ | 119.92 | 122.89 |
| | $C_{14}C_{17}$ | 1.492 | 1.511 | $C_{15}C_{14}C_{18}$ | 118.89 | 120.31 |
| | $C_{14}C_{18}$ | 1.445 | 1.496 | $C_{17}C_{14}C_{18}$ | 107.21 | 113.74 |
| | $C_{15}C_{6}$ | 1.453 | 1.497 | $C_{11}C_{15}C_{14}$ | 120.02 | 123.08 |
| | $C_{15}C_{11}$ | 1.479 | 1.510 | $C_{14}C_{15}C_{6}$ | 119.47 | 120.25 |
| | $C_{14}O_{37}$ | - | 1.494 | $C_{11}C_{15}C_{6}$ | 106.38 | 113.68 |
| | $C_{15}O_{37}$ | - | 1.496 | $C_{14}O_{37}C_{15}$ | - | 57.93 |
| $C_1$ | $C_7C_8$ | 1.464 | 1.497 | $C_8C_7C_{26}$ | 119.24 | 122.30 |
| | $C_7C_{26}$ | 1.453 | 1.542 | $C_{34}C_7C_{26}$ | 104.77 | 110.37 |
| | $C_7C_{34}$ | 1.455 | 1.495 | $C_8C_7C_{34}$ | 120.81 | 123.12 |
| | $C_8C_{10}$ | 1.455 | 1.531 | $C_{10}C_8C_9$ | 117.04 | 122.88 |
| | $C_8C_9$ | 1.445 | 1.498 | $C_{10}C_8C_7$ | 117.20 | 116.80 |
| | $C_7O_{37}$ | - | 1.501 | $C_9C_8C_7$ | 116.36 | 118.53 |
| | $C_8O_{37}$ | - | 1.462 | $C_8O_{37}C_7$ | - | 60.70 |
| $C_s$ | $C_{27}C_{29}$ | 1.480 | 1.500 | $C_{27}C_{29}C_{32}$ | 118.89 | 123.05 |
| | $C_{29}C_{32}$ | 1.464 | 1.508 | $C_{27}C_{29}C_{24}$ | 122.75 | 122.75 |
| | $C_{24}C_{29}$ | 1.434 | 1.519 | $C_{24}C_{29}C_{32}$ | 119.09 | 110.07 |
| | $C_{26}C_{27}$ | 1.460 | 1.527 | $C_{29}C_{27}C_{28}$ | 117.26 | 117.26 |
| | $C_{27}C_{28}$ | 1.434 | 1.515 | $C_{29}C_{27}C_{26}$ | 118.59 | 117.97 |
| | $C_{27}O_{37}$ | - | 1.459 | $C28C_{27}C_{26}$ | 116.17 | 122.94 |
| | $C_{29}O_{37}$ | - | 1.498 | $C_{27}O_{37}C_{29}$ | - | 60.95 |
| $C_2$ | $C_6C_{25}$ | 1.497 | 1.470 | $C_5C_6C_{25}$ | 117.59 | 119.19 |
| | $C_6C_3$ | 1.452 | 1.524 | $C_3C_6C_{25}$ | 118.15 | 122.43 |
| | $C_6C_5$ | 1.449 | 1.459 | $C_3C_6C_5$ | 107.30 | 113.62 |
| | $C_{26}C_{25}$ | 1.454 | 1.459 | $C_{28}C_{25}C_6$ | 118.12 | 122.43 |
| | $C_{28}C_{25}$ | 1.416 | 1.524 | $C_{26}C_{25}C_6$ | 117. 58 | 119.19 |
| | $C_6O_{37}$ | - | 1.477 | $C_{26}C_{25}C_{28}$ | 107.30 | 113.62 |
| | $C_{25}O_{37}$ | - | 1.477 | $C_6O_{37}C_{25}$ | - | 59.69 |
| $D_2$ | $C_4C_8$ | 1.508 | 1.529 | $C_8C_4C_3$ | 118.90 | 118.55 |
| | $C_4C_3$ | 1.441 | 1.513 | $C_8C_4C_6$ | 114.54 | 118.67 |
| | $C_4C_6$ | 1.441 | 1.526 | $C_3C_4C_6$ | 116.47 | 120.95 |
| | $C_8C_7$ | 1.441 | 1.513 | $C_4C_8$ $C_7$ | 118.91 | 118.56 |
| | $C_8C_9$ | 1.441 | 1.526 | $C_4C_8$ $C_9$ | 114.54 | *118.71* |
| | $C_4O_{37}$ | - | 1.448 | $C_7C_8$ $C_9$ | 116.47 | 120.94 |
| | $C_8O_{37}$ | - | 1.447 | $C_4O_{37}C_8$ | - | 63.77 |

To delve deeper into the interaction between the oxygen atom and the carbon cage, comparisons were made of the bond lengths, angles and bond orders of the C-O-C trio of atoms between the most stable endohedral $O@C_{36}$-$D_{3h}$ configurations and the experimental data obtained for Ethylene oxide [42]. The disparity in C-C bond lengths between the $O@C_{36}$-$D_{3h}$ configuration and the Ethylene oxide is minute of order of 0.009 Å, suggesting a non-change in the bond nature remaining simple, whereas the deviation in the C-O-C angle is roughly 3.04 degrees, while the difference in O-C length amounts to approximately 0.069 Å. The C-O binding orders, as per Löwdin and Mayer, stand at 0.938 and 0.751, respectively, with the latter order indicating a slight attenuation in covalent bonding. In Figure 2, the black (blue and red) numerals denote bond lengths and C-O-C angles (Löwdin and Mayer bond orders) within the carbon cage, respectively.

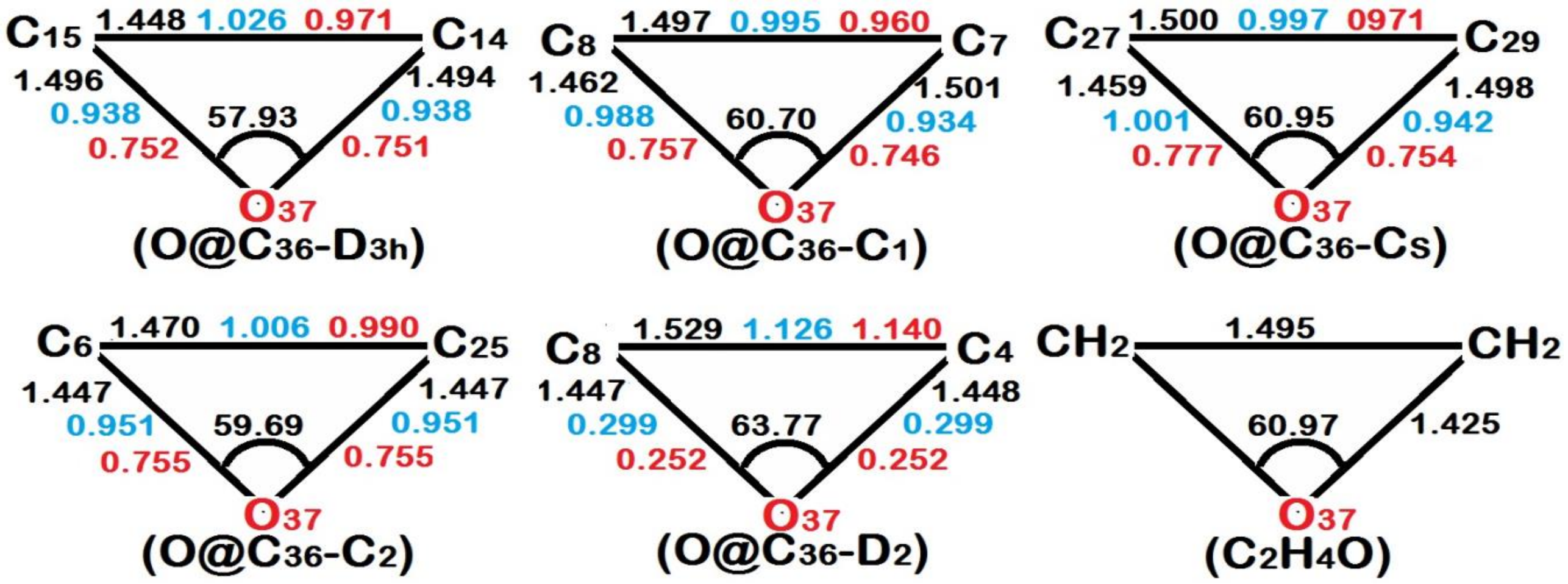

Fig. 2. illustrates bond lengths and angles (depicted in black) as well as bond orders (represented by Löwdin in blue and Mayer in red) for the five $O@C_{36}$ isomers, obtained through DFT/B3LYP/6-31G* calculations. Additionally, experimental bond lengths and angles for the reference molecule $C_2H_4O$ are provided for comparison.

Table 2. Electronic energies of $C_{36}$ isomers and their endohedrals, the relative energies (computed as energy differences from the most stable configuration), the encapsulating energies $E_{en}$, the Atomization energies AE and the number of pentagon adjacencies, e55, using the DFT method at B3LYP/6-31G* level. We adopted the values -2039.693 (-1027.3) eV for the Oxygen $E_O$ (Carbon $E_C$) atom energy.

| Energy (eV) | | | | | Relative energy (eV) | | | Atomization energy (eV/atom) | |
|---|---|---|---|---|---|---|---|---|---|
| | | | | | $C_{36}$ | | | | |
| Sym. | e55 | $C_{36}$ | $O@C_{36}$ | $E_{en}$ | Our work | Other work [19] | $O@C_{36}$ | $C_{36}$ | $O@C_{36}$ |
| $D_{3h}$ | 15 | -37289.617 | -39330.887 | -1.580 | 0.000 | 0.000 | 0.000 | 8.521 | 0.043 |
| $C_1$ | 15 | -37288.761 | -39330.577 | -2.126 | 0.856 | 0.983 | 0.310 | 8.497 | 0.057 |
| $C_s$ | 16 | -37287.922 | -39329.727 | -2.115 | 1.695 | 1.801 | 3.431 | 8.474 | 0.057 |
| $C_2$ | 16 | -37287.879 | -39328.824 | -1.255 | 1.738 | 2.085 | 4.627 | 8.473 | 0.034 |
| $D_2$ | 18 | -37286.094 | -39328.215 | -2.431 | 3.523 | 3.009 | 6.102 | 8.423 | 0.066 |

Table 2 illustrates the energy values corresponding to various fullerene isomers and their endohedral counterparts, alongside their relative energies defined as differences from the most stable configuration ($D_{3h}$ and $O@C_{36}$-$D_{3h}$ for fullerenes and their endohedrals, respectively). It shows also the encapsulation ($E_{en}$) and Atomization (AE) energies, which are defined as follows.

$$E_{en} = E(O@C_{36}) – [E(C_{36}) + E(O)] \quad (1)$$

$$AE(C_{36}) = \frac{36.E_{Carbon} - E_{C_{36}}}{36} \quad (2)$$

$$AE(O@C_{36}) = \frac{[E(C_{36}) + E(O)] - E(O@C_{36})}{37} \quad (3)$$

The negative values of $E_{en}$ means that the oxygen encapsulation represents an energetically favorable process in that its stabilization within the $C_{36}$ fullerene is a systematically exothermic reaction. We see that the energy ordering of these compounds follows the pentagonal adjacency penalty rule (PAPR), indicating that stability increases as the number of bonds shared between pentagons decreases, reducing structural stress. Actually, when pentagons share edges, this leads to significant stress and instability in the carbon cage structure, as the carbon atoms are forced into less favorable bond angles, resulting in higher reactivity and less stable configurations. Notably, the energetic stability hierarchy persists following the incorporation of the oxygen atom.

Moreover, while the most stable isomer exhibits the highest atomization energy, incorporation of the oxygen atom results in only a significant corresponding decrease for the $C_{36}$-$D_{3h}$, $C_{36}$-$C_1$, $C_{36}$-$C_s$, $C_{36}$-$C_2$ and $C_{36}$-$D_2$ isomers by 198.54%, 146.88%, 147.24%, 248.88% and 127.20%, respectively.

### 3. 2. Electronic and electric properties

#### 3. 2. 1. The HOMO/LUMO energies, energy gap and Fermi energy

Figure 3 illustrates the $E_{HOMO}$ (Highest occupied molecular orbital), $E_{LUMO}$ (Lowest unoccupied molecular orbital), Egap (energy gap) and $E_F$ (Fermi energy, i.e. the energy required for adding an electron to the system) values for both fullerene $C_{36}$ and its endohedral variant $O@C_{36}$. These values serve to characterize the electronic properties of the molecules, including their electron donating or accepting abilities. We adopt for the $E_F$ and Egap values for both fullerene $C_{36}$ isomers and its endohedral $O@C_{36}$ the following defining equations:

$$E_F = (E_{HOMO} + E_{LUMO})/2 \quad (4)$$

$$E_{gap} = E_{LUMO} - E_{HOMO} \quad (5)$$

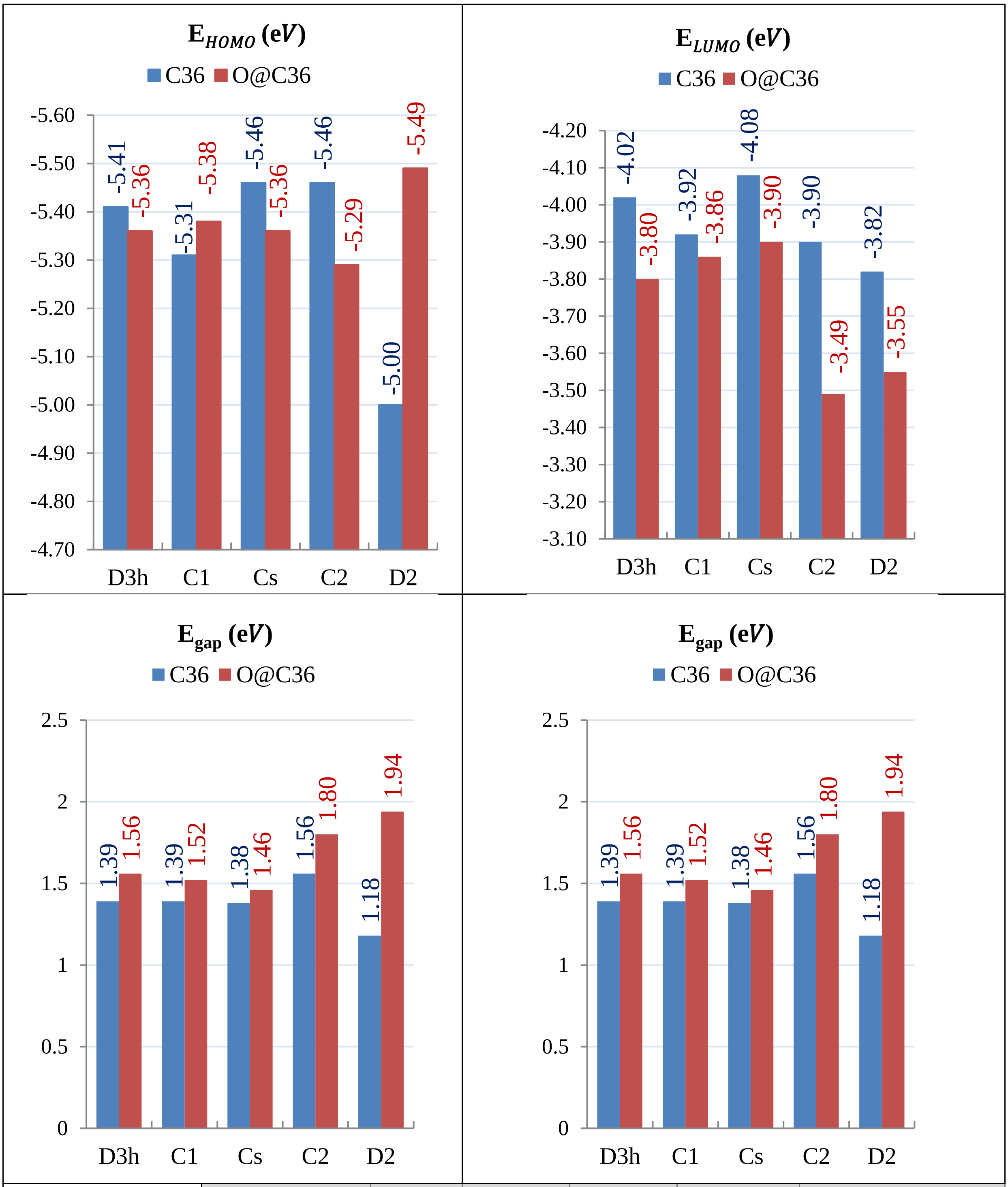

| Previous works on $C_{36}$ isomers | Sym. | $D_{3h}$ | $C_1$ | $C_s$ | $C_2$ | $D_2$ |
|---|---|---|---|---|---|---|
| | $E_F$ [20] | -5.08 | -5.03 | -5.05 | -5.00 | -5.01 |
| | Egap [19] | 0.45 | 0.33 | 0.38 | 0.20 | 0.63 |
| | Egap [20] | 0.48 | 0.50 | 0.57 | 0.48 | 0.60 |

Fig. 3. The $E_{HOMO}$, $E_{LUMO}$, $E_{gap}$ and Fermi energies (in eV) for $C_{36}$ isomers and their endohedral counterparts using the DFT method at B3LYP/6-31G* level.

The findings illustrate that the $C_{36}$-$D_2$ ($C_{36}$-$C_2$) isomer displays the highest HOMO and LUMO energies before (after) encapsulation. Conversely, the $C_{36}$-$C_s$ and $C_{36}$-$C_2$ ($C_{36}$-$D_2$) isomers exhibit the lowest HOMO energy before (after) encapsulation, while the $C_{36}$-$C_s$ isomer presenting the lowest LUMO energy both before and after encapsulation. The HOMO energy decreases due to oxygen atom encapsulation for the $C_{36}$-$D_{3h}$, $C_{36}$-$C_s$, and $C_{36}$-$C_2$ isomers by 0.92%, 1.83% and 3.11%, respectively. In contrast, for the $C_{36}$-$C_1$ and $C_{36}$-$D_2$ isomers, there is an increase of 1.32% and 9.80%, respectively, compared to their states prior to encapsulation. Additionally, the LUMO energy increases due to oxygen atom encapsulation for the $C_{36}$-$D_{3h}$, $C_{36}$-$C_1$, $C_{36}$-$C_s$, $C_{36}$-$C_2$ and $C_{36}$-$D_2$ isomers by 5.47%, 1.53%, 4.41%, 10.51% and 7.07%, respectively, compared to their previous states before encapsulation.

Furthermore, a significant increase in energy gap is observed following oxygen atom encapsulation for the $C_{36}$-$D_{3h}$, $C_{36}$-$C_1$, $C_{36}$-$C_s$, $C_{36}$-$C_2$ and $C_{36}$-$D_2$ isomers by 12.23%, 9.35%, 5.8%, 15.38% and 64.41%, respectively, compared to their initial states prior to encapsulation. Notably, the $C_{36}$-$C_2$ isomer possesses the widest energy gap ($E_{gap}$= 1.56 eV), while the $C_{36}$-$D_2$ isomer exhibits the narrowest one ($E_{gap}$= 1.18 eV). Post-encapsulation, the energy gap further widens for the $C_{36}$-$D_2$ isomer ($E_{gap}$= 1.94 eV), while it narrows for the $C_{36}$-$C_s$ isomer ($E_{gap}$= 1.46 eV). Despite previous studies highlighting semiconductor behavior in $C_{36}$ isomers, the energy gap values obtained in this research significantly exceed previously reported ones, as stated in Figure 3.

Additionally, the Fermi energy ($E_F$), representing the average energy of LUMO and HOMO, is analyzed. The results demonstrate that the $C_{36}$-$D_2$ isomer has the highest Fermi energy ($E_F$ = -4.41eV), while the $C_{36}$-$C_s$ isomer has the lowest Fermi energy ($E_F$ = -4.77 eV). Post-encapsulation, the $C_{36}$-$C_2$ isomer exhibits the highest Fermi energy ($E_F$ = -4.39 eV), while the $C_{36}$-$C_s$ isomer demonstrates the lowest one ($E_F$ = -4.63 eV). Additionally, the results reveal an increase (decrease) in Fermi energy due to oxygen atom encapsulation for the $C_{36}$-$D_{3h}$, $C_{36}$-$C_s$ and $C_{36}$-$C_2$ ($C_{36}$-$D_2$) isomers by 2.97%, 2.94% and 6.20% (2.49%), respectively, compared to their previous states before encapsulation.

### 3. 2. 2. Potential ionization, affinity, hardness, chemical potential, softness and electrophilicity

We discuss now the energy quantities of potential ionization (I), affinity (A), global hardness($\eta$), chemical potential ($\mu$ *or* χ), Softness (S) and global electrophilicity (ω) for fullerene isomers and their endohedral O@$C_{36}$. $I$ ($A$) is the amount of energy taken (released) when removing (adding) an electron from (to) a molecule, computed as the energy difference between the cation (anion) and the neutral molecule:

$$I = E_{\text{optimized cation}} - E_{\text{optimized neutral}} \quad (6)$$

$$A = E_{\text{optimized neutral}} - E_{\text{optimized anion}} \quad (7)$$

$I$ and $A$ allow to compute $\mu$ (χ) measuring the ability of a system to donate or accept electrons with negative (positive) values for all our isomers, indicating that they all tend to gain electrons.

As to $\eta$, defined as the energy required to add or remove an electron from the molecule, it decreases after encapsulation, indicating that the fullerene cage becomes more susceptible to electron transfer. Its half inverse, S, increases after encapsulation, indicating a decrease in the stability of the fullerene cage. Finally, $\omega$ measures the ability of a system to accept electrons with positive values for all isomers under study, indicating that they all are good electron acceptors. These energy quantities are defined as [43]:

$$\chi = (I + A)/2 \quad (8)$$

$$\mu = -(I + A)/2 \quad (9)$$

$$\eta = (I - A)/2 \quad (10)$$

$$S = 1/2\eta \quad (11)$$

$$\omega = \mu^2/2\eta \quad (12)$$

Table 3. Potential ionization ($I$), affinity ($A$), global hardness ($\eta$), chemical potential ($\mu$), softness ($S$) and global electrophilicity ($\omega$), in eV, of $C_{36}$ and $O@C_{36}$ using the DFT method at B3LYP/6-31G* level.

| Sym. | | $D_{3h}$ | $C_1$ | Cs | $C_2$ | $D_2$ |
|---|---|---|---|---|---|---|
| $C_{36}$ | $E_{C36}$ | -37289.617 | -37288.761 | -37287.922 | -37287.879 | -37286.094 |
| | $E_{C36^+}$ | -37282.865 | -37282.192 | -37281.310 | -37281.118 | -37279.952 |
| | $E_{C36^-}$ | -37292.318 | -37291.519 | -37290.733 | -37290.489 | -37288.905 |
| | I | 6.752 | 6.569 | 6.612 | 6.761 | 6.142 |
| | A | 2.701 | 2.758 | 2.811 | 2.610 | 2.811 |
| | $\eta$ | 2.026 | 1.906 | 1.901 | 2.075 | 1.665 |
| | S | 0.24685 | 0.26240 | 0.26309 | 0.24091 | 0.30021 |
| | $\mu$ | -4.727 | -4.664 | -4.711 | -4.685 | -4.476 |
| | χ | 4.727 | 4.664 | 4.711 | 4.685 | 4.476 |
| | ω | 5.515 | 5.707 | 5.840 | 5.289 | 6.016 |
| $O@C_{36}$ | $E_{O@C36}$ | -39330.887 | -39330.577 | -39329.727 | -39328.824 | -39328.215 |
| | $E_{O@C36^+}$ | -39324.263 | -39322.992 | -39322.066 | -39322.607 | -39321.467 |
| | $E_{O@C36^-}$ | -39333.470 | -39331.893 | -39331.414 | -39330.919 | -39330.529 |
| | I | 6.624 | 7.585 | 7.661 | 6.217 | 6.748 |
| | A | 2.583 | 1.316 | 1.687 | 2.095 | 2.314 |
| | $\eta$ | 2.021 | 3.135 | 2.987 | 2.061 | 2.217 |
| | S | 0.24746 | 0.15952 | 0.16739 | 0.24260 | 0.22553 |
| | $\mu$ | -4.6035 | -4.4505 | -4.674 | -4.156 | -4.531 |
| | χ | 4.6035 | 4.4505 | 4.674 | 4.156 | 4.531 |
| | ω | 5.244299 | 3.159507 | 3.656893 | 4.19028 | 4.630 |

Table 3 presents the values for these energy quantities for $C_{36}$ fullerene isomers ($D_{3h}$, $C_1$, $C_s$, $C_2$, $D_2$) and their endohedral complexes after encapsulation with oxygen atoms, derived from DFT calculations at B3LYP/6-31G* level. We see that the $D_2$ isomer exhibits the lowest potential ionization value (I = 6.142 eV), making it the superior electron donor, and that the $C_s$ isomer shows the highest affinity value (A = 2.811 eV), rendering it the most efficient electron acceptor among all the isomers. The chemical hardness value of the $D_2$ isomer (η = 1.665 eV) is the lowest among all the isomers, indicating its heightened reactivity compared to others, and so it is identified as the most reactive isomer. Additionally, the $D_{3h}$ isomer showcases a higher electronegativity value (χ = 4.727 eV) than any other isomer, while its electrophilic index (ω = 6.016 eV) suggests it as the strongest electrophile among all isomers. Moreover, the potential ionization and global hardness values for $O@C_{36}$-$C_1$, $O@C_{36}$-$C_s$ and $O@C_{36}$-$D_2$ isomers surpass those of the original molecules, whereas the electron affinity and electrophilic index values for $O@C_{36}$ isomers are lower than the original ones. Similarly, the electronegativity values decreased for all isomers except for the $O@C_{36}$-$D_2$ isomer, which showed an increase. Furthermore, fullerene isomers ($D_2$, $C_2$) demonstrate greater energy stabilization when acquiring an additional electronic charge from the surroundings, displaying the least tendency to accept more electrons compared to other fullerene isomers. Before oxygen atom encapsulation, the $C_s$ isomer shows the greatest electron exchange, while after encapsulation, it is the $C_2$ isomer.

### 3. 2. 3. Atomic charge

There are two common notions of atomic charges. The atomic Mulliken charge represents the charge an atom would possess if its electron density were isolated, whereas the atomic Löwdin charge reflects the charge based on the molecule's electrostatic potential. Table 4 exhibits values of these charge values for each isomer before and after the encapsulation and for the Oxygen and the two connected Carbon atoms. Notably, the oxygen atom exhibits a negative charge across all endohedral complexes, while the charges of carbon atoms in the fullerene isomers vary based on their positions and interactions with the encapsulated oxygen atom.

Table 4. Mulliken and Löwdin charge analysis, in units of the elementary charge e, of $C_{36}$ isomers and theirs endohedral using the DFT method at B3LYP/6-31G* level.

| Sym. | Atom Label | $C_{36}$ | | O@$C_{36}$ | |
|---|---|---|---|---|---|
| | | Mulliken charge | Löwdin charge | Mulliken Charge | Löwdin charge |
| $D_{3h}$ | $C_{14}$ | 0.015945 | -0.001694 | 0.123412 | 0.100804 |
| | $C_{15}$ | 0.019809 | 0.001864 | 0.12314 | 0.101089 |
| | $O_{37}$ | | | -0.481815 | -0.25092 |
| $C_1$ | $C_7$ | 0.008438 | -0.011886 | 0.120283 | 0.119822 |
| | $C_8$ | 0.029546 | -0.003747 | 0.122212 | 0.090724 |
| | $O_{37}$ | | | -0.445202 | -0.217389 |
| Cs | $C_{27}$ | 0.040971 | -0.008105 | 0.111280 | 0.089255 |
| | $C_{29}$ | 0.007669 | 0.000706 | 0.123549 | 0.119134 |
| | $O_{37}$ | | | -0.450121 | -0.223721 |
| $C_2$ | $C_6$ | 0.020066 | -0.002614 | 0.105208 | 0.096585 |
| | $C_{25}$ | 0.019957 | -0.002493 | 0.105202 | 0.096586 |
| | $O_{37}$ | | | -0.017599 | -0.222329 |
| $D_2$ | $C_4$ | 0.026702 | -0.000264 | 0.103582 | 0.095223 |
| | $C_8$ | 0.026704 | -0.000262 | 0.103831 | 0.095201 |
| | $O_{37}$ | | | -0.440843 | -0.20587 |

Following the encapsulation, the Mulliken charges remained positive and exhibited an increase for the two carbon atoms. On the other hand, the negative (positive) Löwdin charges transitioned to (remained) positive charges for these carbon atoms. In all cases, there was an augmentation in the positive charge values for the $D_{3h}$ and $C_s$ isomers. Overall, the Löwdin atomic charge values are notably smaller (in absolute value) than the Mulliken charge values, but all charge values for the two Carbon atoms approach each other after encapsulation.

### 3. 2.4. Electric dipole moments and polarizabilities

Again, we use the DFT method at B3LYP/6-31G* level in order to compute the theoretical dipole moment and polarizability of the $C_{36}$ fullerene isomers and their respective endohedral complexes with oxygen atoms. Generally, the dipole moment and polarizability values of the endohedral complexes surpass those of the corresponding $C_{36}$ fullerene isomers, with the exception of the $D_{3h}$ isomer. These computed values offer insights into the electric and optical characteristics of the molecules, which hold significance across various applications.

Table 5. Electric dipole moments (debye), molecular polarizabilities and atomic polarizabilities of $C_{36}$ isomers and their endohedral using the DFT method at B3LYP/6-31G* level.

| | Sym. | $D_{3h}$ | $C_1$ | $C_s$ | $C_2$ | $D_2$ |
|---|---|---|---|---|---|---|
| p (Debye) | $C_{36}$ | 0.079 | 0.998 | 0.494 | 0.465 | 0.000 |
| | $O@C_{36}$ | 1.011 | 1.011 | 1.091 | 1.352 | 0.320 |
| Molecular polarizability (a.u) | $C_{36}$ | 296.89 | 289.59 | 295.27 | 284.42 | 296.16 |
| | $O@C_{36}$ | 293.35 | 290.01 | 296.76 | 291.03 | 304.82 |

| Atomic polarizabilities (a.u) | | | | | |
|---|---|---|---|---|---|
| | Previous works [44] | | This work | | |
| | Value | Comments | Value | $36\alpha_C$ | $36\alpha_C + \alpha_O$ |
| $\alpha_C$ | 11.39 | NR, CASPT2, ML res. | $\alpha_C$= 5.46978 | 196.91 | 198.98 |
| | 11.67 ± 0.07 | NR, CCSD(T), ML res. | | | |
| | 11.26 ± 0.20 | R, Dirac+Gaunt, CCSD(T) | | | |
| | 11.63 | NR, CCSD (T) | | | |
| | 11.3 ± 0.2 | Recommended | | | |
| $\alpha_O$ | 5.41 ± 0.11 | NR, PNO-CEPA, ML res. | $\alpha_O$= 2.06731 | | |
| | 5.4 ± 0.7 | NR, CASPT2, ML res. | | | |
| | 5.24 ± 0.04 | NR, CCSD(T), ML res. | | | |
| | 5.15 | NR, CCSD(T) | | | |
| | 5.2 ±0.4 | exp. | | | |
| | 5.3 ± 0.2 | Recommended | | | |

Table 5 reveals a discrepancy with the energy stability order of the compounds. Notably, the $C_1$ and $D_2$ isomers exhibited respectively the highest and lowest dipole moment values, whereas, for post-encapsulation, it is the $C_2$ and $D_2$ isomers which assumed these positions. Furthermore, encapsulation of the oxygen atom led in general to increased dipole moment values, with the most significant changes observed in the $D_{3h}$ and $C_2$ isomers, indicating a decrease in molecular symmetry

Despite the introduction of asymmetric and electronegativity effects, as well as changes in atomic charge density, the molecular polarizability value experienced only a slight increase, ranging from 0.4% to 0.5% post-encapsulation.

Regarding atomic polarizabilities, for free Carbon and Oxygen atoms, it is observed that DFT values significantly underestimate those obtained from more advanced methodologies as stated in previous studies [44]. Furthermore, the summation of atomic polarizabilities tends to underestimated values when compared to molecular ones.

### 3.3. Thermodynamic analysis

We provide here computed values of the Zero-point energy (ZPE) and the inner energy for various $C_{36}$ fullerene isomers, including their endohedral forms, and display the calculated corresponding values for enthalpy, entropy and Gibbs energy at a temperature of 298.15 K.

Table 6. Zero-point energy (ZPE) and inner energy of $C_{36}$ isomers and their endohedral isomers using the DFT method at B3LYP/6-31G* level.

| Sym. | Zero-point energy ZPE (kcal/mol) | | Inner energy (kcal/mol) | |
|---|---|---|---|---|
| | $C_{36}$ | $O@C_{36}$ | $C_{36}$ | $O@C_{36}$ |
| $D_{3h}$ | 136.4828 | 135.9243 | -859772.9917 | -906843.6165 |
| $C_1$ | 134.6254 | 135.9243 | -859753.3445 | -906836.8457 |
| Cs | 134.0795 | 135.6608 | -859734.4628 | -906817.5059 |
| $C_2$ | 134.6129 | 136.0373 | -859733.0697 | -906796.4907 |
| $D_2$ | 133.3390 | 135.4788 | -859692.9155 | -906782.8738 |

Table 6 indicates that the ranking of the Zero-point energy (ZPE) values before and after encapsulation for the five $C_{36}$ isomers does not correspond to their relative energy stabilities stated in Table 2 in that the $C_2$ isomer deviates from its stability ordering. Prior to encapsulation, the $D_{3h}$ ($D_2$) isomer has the highest (lowest) ZPE. Inner energy values before and after encapsulation perfectly follow the stability order of the isomers.

Figure 4. shows other thermodynamic quantities comprising the enthalpy, entropy and Gibbs energy for the isomers under study. We see that the enthalpy values before and after encapsulation perfectly follow the stability order of the isomers, and so -regarding thermal stability- the $D_{3h}$ isomer is the most stable. Again, the stated values of enthalpy indicate that the oxygen encapsulation represents generically an exothermic process in that its values for the $O@C_{36}$-system are always algebraically beneath, albeit slightly, the sum of the $C_{36}$-system enthalpy and that of a single oxygen atom (-2040 eV).

The data also indicates that the encapsulation of oxygen results in a small increase in entropy, ranging from 1.2% ($D_{3h}$) to 3.3% ($C_1$), compared to the $C_{36}$ isomers before encapsulation. Moreover, encapsulation does not affect the energy classification based on Gibbs energy, but rather enhances the stability of the compounds.

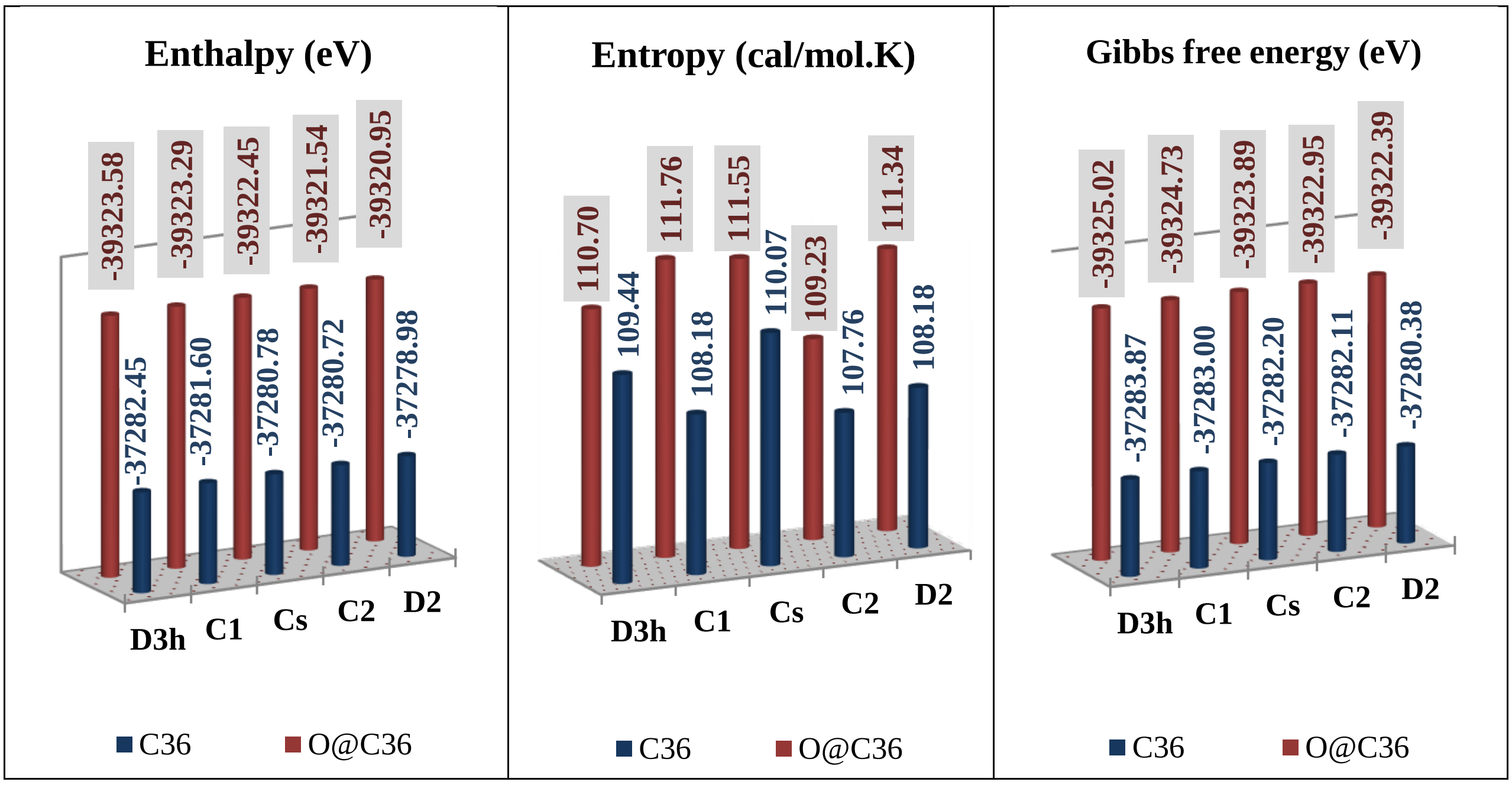

Fig. 4. The enthalpy, entropy and Gibbs energy values of $C_{36}$ isomers and their endohedral counterparts using the DFT method at B3LYP/6-31G* level.

### 3.4. Vibration frequencies and infrared spectra

Figure 5 illustrates the calculated vibration frequencies, evaluated in $cm^{-1}$, alongside the highest IR absorption intensities for both $C_{36}$ and $O@C_{36}$ fullerene isomers.

We observe here that the highest absorption intensity for $C_1$ corresponds to a frequency of around half the corresponding frequency in the $D_{3h}$, $C_s$, $C_2$ and $D_2$ symmetries before oxygen encapsulation. However, after encapsulation, this frequency increases (decreases) by 86.4% for $C_1$ and 6.8% for $C_s$ (8.4% for $C_2$ and 26.2% for $D_2$). It is worth noting also that the absorption intensity values decrease (increase) after encapsulation by 38.4% for $D_{3h}$ and 50.7% for $D_2$ (39.2% for $C_1$, 104.6% for Cs and 84% for $C_2$) compared to their pre-encapsulation values. The vibration pattern accompanying a larger change in the dipole moment during the vibration process corresponds to a higher absorption intensity.

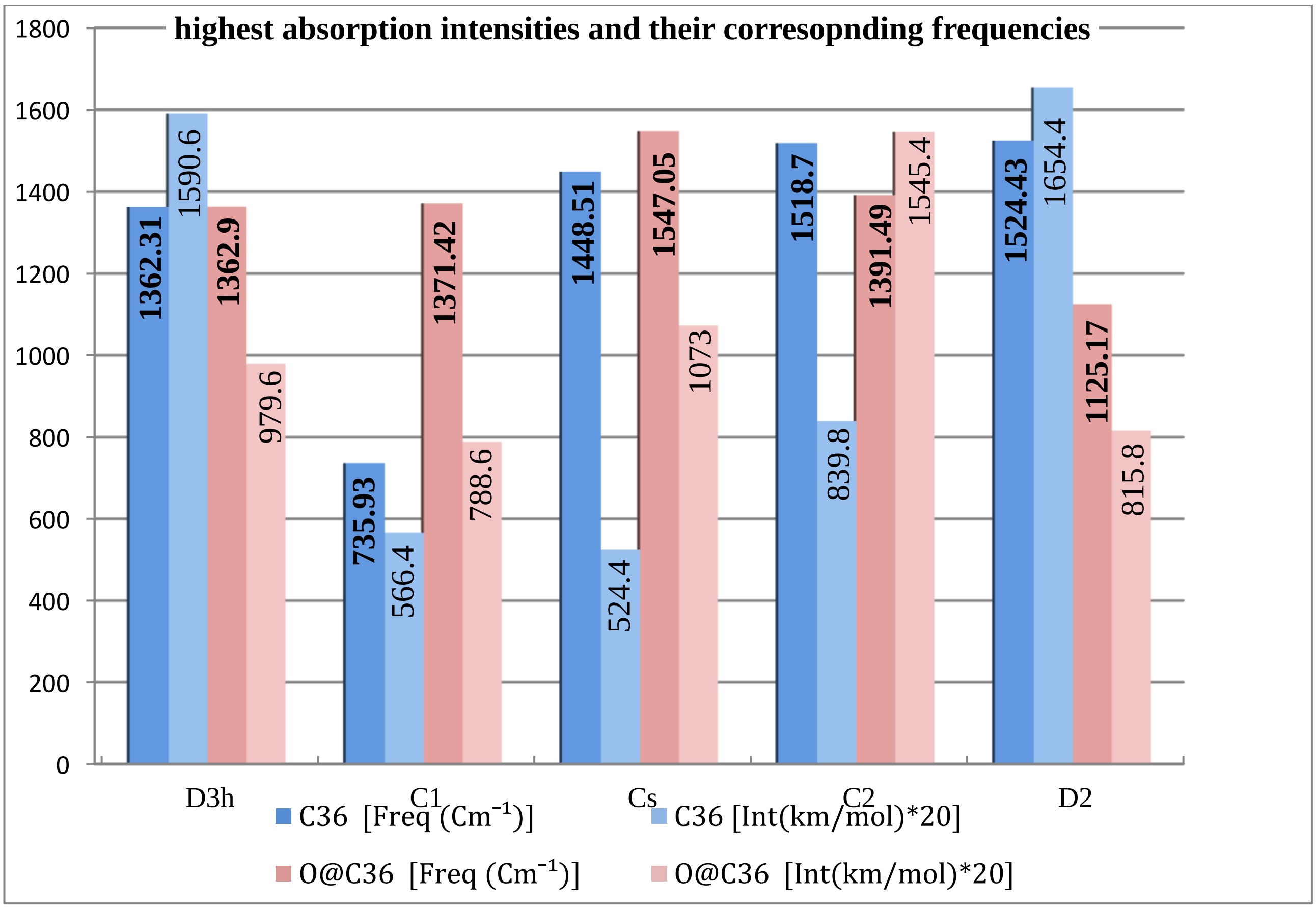

Fig.5. Vibration frequencies ($cm^{-1}$) and IR absorption intensity for $C_{36}$ and $O@C_{36}$ fullerene isomers, using the DFT method at B3LYP/6-31G* level. For illustration purposes, an amplification factor equaling 20 was used in IR intensity.

### 3.5. NMR spectra for the $C_{36}$ and $O@C_{36}$ fullerene isomers

Table 7 provides the isotropic chemical shielding values for both $C_{36}$ and $O@C_{36}$ fullerene isomer. We observe that the NMR chemical shifts for the carbon atoms adjacent to the encapsulated oxygen atom have increased significantly after the oxygen atom was placed inside the $C_{36}$ fullerene isomers, when compared to their values before encapsulation

Table 7. NMR chemical shielding (ppm) for $C_{36}$ and O@$C_{36}$ isomers using the DFT method at B3LYP/6-31G* level.

| Sym. | Atom | $C_{36}$ | O@$C_{36}$ | Sym. | Atom | $C_{36}$ | O@$C_{36}$ |
|---|---|---|---|---|---|---|---|
| $D_{3h}$ | $C_{14}$ | 46.21 | 109.24 | $C_2$ | $C_6$ | 52.87 | 96.13 |
| | $C_{15}$ | 35.15 | 109.201 | | $C_{25}$ | 52.96 | 96.13 |
| | $O_{37}$ | | -245.01 | | $O_{37}$ | | -9.99 |
| $C_1$ | $C_7$ | 36.18 | 101.83 | $D_2$ | $C_4$ | 54.56 | 106. 89 |
| | $C_8$ | 52.48 | 108.93 | | $C_8$ | 54.56 | 106.85 |
| | $O_{37}$ | | 53.049 | | $O_{37}$ | | 420.32 |
| Cs | $C_{26}$ | 54.44 | 44.04 | | | | |
| | $C_{28}$ | 23.89 | 58.75 | | | | |
| | $O_{37}$ | | 38.35 | | | | |

## 4. Conclusions

In this study, we conducted a comprehensive investigation into various aspects of $C_{36}$ and its O@$C_{36}$ isomers ($D_{3h}$, $C_1$, $C_s$, $C_2$ and $D_2$) utilizing the DFT method at B3LYP/6-31G*level. Our analysis covered geometry optimizations, relative stability, atomization energies, static electric polarizabilities, energy gap, potential ionization, affinity, hardness, chemical potential, softness, electrophilicity, and thermodynamic parameters. Additionally, we examined vibration frequencies, as well as infrared and NMR spectra. Consistently, our findings revealed that the $D_{3h}$ isomer emerged as the most stable among the studied $C_{36}$ isomers, while the $D_2$ isomer displayed the least stability. This hierarchy remained unchanged even after Oxygen encapsulation. It is worth noting that geometrically the oxygen atom did not occupy the central position within the fullerene cage

Regarding the energy gap, the $D_2$ isomer exhibited the smallest gap (1.18 eV) compared to other isomers, while it showed the largest gap (1.94 eV) after encapsulation. Moreover, our results demonstrated an increase (or decrease) in Fermi energy due to oxygen atom encapsulation for specific isomers, such as $C_{36}$-$D_{3h}$, $C_{36}$-$C_s$, and $C_{36}$-$C_2$ ($C_{36}$-$D_2$), by 2.97%, 2.94%, and 6.20% (2.49%) respectively, compared to their pre-encapsulation states. Furthermore, the encapsulation of oxygen atoms induced changes in the potential ionization and hardness values among the $C_1$, $C_s$ and $D_2$ ($D_{3h,}$ $C_2$) isomers, with a distinct increase (decrease). Conversely, a marginal decrease in Affinity, electronegativity and electrophilicity values was observed across all isomers subsequent to encapsulation.

Moreover, our study examined alterations in IR and absorption intensity induced by encapsulation, revealing a reduction (augmentation) in IR absorption intensity for $D_{3h}$, $C_2$, and $D_2$ ($C_1$ and $C_s$) isomers. Additionally, we delved into properties such as polarizability and atomic charge as part of our analysis.

Crucially, our investigation into BSSE corrections demonstrated that their influence on both geometric and electronic properties of these rigid fullerene systems is minimal. Comparative calculations with and without BSSE corrections revealed negligible differences in bond lengths, angles, total energies, and atomic charge distributions. For example, variations in bond lengths and valence angles between corrected and uncorrected methods were less than 0.4%, and atomic charges on the oxygen atom changed by less than 0.3%. These results confirm that the B3LYP/6-31G* approach provides robust and reliable predictions for large, rigid fullerenes and their endohedral derivatives, even without explicit BSSE corrections. Thus, for similar systems where computational efficiency is a priority, BSSE corrections may be safely omitted without compromising accuracy.

In summary, our findings not only clarify the structural and electronic behavior of $C_{36}$ and $O@C_{36}$ isomers but also validate the computational methodology for future studies of large, rigid fullerenes and their derivatives.

We hope the results of this work would incite/help experimentalists to fabricate/characterize the investigated fullerene systems, and to interpret their spectra, with some potential medical (technological) applications, such as in drug delivery (quantum computing), making full use of the amalgamation of a small size oxygen atom with a fullerene system acting as a Faraday cage.

**Data Availability Statement:** All data generated or analyzed during this study are included in this published Article.
**Conflicts of Interest:** The authors declare no conflict of interest.

**Author Contributions Statement:** All authors contributed equally to this work.

**Acknowledgement:** We thank Alaa Alddin Mardini from MIPT for help in this analysis. N.C. acknowledges support from the CAS PIFI fellowship, and from the Humboldt Foundation.

**Appendix:**

In this appendix, we outline some technical points addressed in our work. Section 1 revisits BSSE general concepts, whereas section 2 state results including/excluding the BSSE corrections for the studied isomers with $D_{3h}$ symmetry, justifying non-consideration of BSSE in the work. Section 3, deals with convergence issues in ORCA.

## 1) Revisiting the Basis Set Superposition Error (BSSE): Definition and Context/Use in the Work

This section outlines the Basis Set Superposition Error (BSSE) issue, detailing its conceptual framework, methodological basis, correction techniques (with constraints), practical scope, computational demands, structural flexibility considerations, and ends with an assessment of B3LYP/6-31G*'s accuracy-efficiency trade-off in this work.

### 1.1) Definition-Problem

BSSE is a well-known issue in computational quantum chemistry that arises when using finite, atom-centered basis sets to approximate molecular wavefunctions. The BSSE introduces systematic inaccuracies in quantum chemical calculations by artificially stabilizing molecular systems due to the overlap of finite basis sets between fragments. When two or more molecules (or fragments of a molecule) interact, their basis functions overlap, allowing each fragment to "borrow" basis functions from the other. This artificial increase in the effective basis set lowers the system's calculated energy, leading to an overestimation of binding energies and other derived properties.

### 1.2) BSSE Correction Methods

To address the BSSE problem, two primary correction methods are used:

- **Counterpoise (CP) Method:**
  The most widely used approach for correcting BSSE is the counterpoise method, introduced by Boys and Bernardi. In this method, the energy of each monomer is recalculated in the presence of the basis functions of the other monomer(s), but without their nuclei or electrons (these are called "ghost orbitals"). The corrected interaction energy is then obtained by subtracting the energies of the monomers (computed with ghost orbitals) from the energy of the full complex. This correction is applied *a posteriori* and aims to remove the artificial stabilization caused by basis set overlap.
- **Chemical Hamiltonian Approach (CHA):**
  Alternatively, the CHA modifies the Hamiltonian itself to prevent basis set mixing *a priori*. This is done by removing terms in the Hamiltonian that would allow basis functions from different fragments to interact. While conceptually different, CHA and CP often yield similar results, especially for intermolecular interaction energies.

### 1.3) Basis Set size Dependency, applicability, limitations, and Correction Trade-offs

Literature-based practical insights identify three key factors influencing BSSE:

- Basis Set Size: BSSE decreases as the basis set size increases and becomes negligible in the limit of a complete (infinite) basis set.
- BSSE decreases with larger basis set size (e.g., cc-pVQZ), but residual Basis Set Incompleteness Error (BSIE) persists, and becomes negligible in the limit of a complete (infinite) basis set. For example, CP corrections become negligible (<0.01 kcal/mol) in post-CCSD(T) calculations with haVQZ.
- Overcorrection risks: CP may over-stabilize systems in hydrogen-bonded clusters or when using very small basis sets (e.g., cc-pVDZ without diffuse functions).
- Applicability: BSSE is particularly significant in systems dominated by weak, non-covalent interactions (e.g., hydrogen bonding, van der Waals complexes) and can also affect intramolecular interactions.
- Limitations: Both CP and CHA have limitations. For example, CP corrections can sometimes overcorrect, especially in small basis sets or in regions of the potential energy surface where the effect of ghost orbitals is inconsistent.

### 1.4) Molecule classes where BSSE Corrections are critical

BSSE corrections are significant for certain molecules and systems where basis set overlap can significantly distort calculated properties. BSSE corrections are crucial for:

- Weakly bound and noncovalent complexes
- Molecular clusters and crystals
- Large, flexible, or intramolecularly interacting molecules
- Anionic and diffuse systems

Neglecting BSSE in these cases can lead to significant errors in calculated energies, geometries, and other molecular properties. We describe briefly each case:

- Noncovalent interactions: such as in hydrogen bonding, π-stacking, and van der Waals forces
- In weakly interacting systems-including Van der Waals complexes, hydrogen-bonded structures, and dispersion-dominated assemblies, BSSE strongly influences systems such as helium/argon dimers and molecular complexes like ArHF, $HCO^-$, and $(HF)_2$. In these cases, uncorrected calculations often yield exaggerated binding energies and irregular convergence of properties like dissociation energies and equilibrium bond lengths. The CP correction dramatically improves accuracy for these systems.
- Molecular Clusters and Crystals: For molecular crystals and clusters (e.g., benzene, aspirin, oxalyl dihydrazide), BSSE can accumulate with cluster size, leading to significant errors in cohesive and interaction energies. CP correction is essential for an accurate description, especially in larger aggregates.
- Strongly Bound Diatomics: Even in strongly bound diatomics like $N_2$, HF, and HCl, BSSE can impact the convergence of structural and vibrational properties (such as bond lengths and harmonic frequencies), though the effect is less dramatic than in weakly bound systems. CP correction often still improves the reliability of computed properties.
- Intramolecular effects, Conformational Energies and Noncovalent Interactions: BSSE is not limited to intermolecular interactions. In large, flexible molecules or those with significant intramolecular noncovalent interactions (e.g., aromatic rings, helicenes, peptides), different parts of the molecule can "borrow" basis functions from each other, leading to artificial stabilization and geometric distortions. Intramolecular CP corrections are particularly important in such cases to avoid artifacts like artificial puckering in benzene or misestimated conformational energies in peptides.
- Anions and Systems with Diffuse Electron Density: Systems such as anions, which have more diffuse electron densities, are especially sensitive to BSSE. Failure to correct for BSSE in these systems can result in unusually large binding energies and other unphysical effects.

## 1.5) Efficiency and Limitations of B3LYP/6-31G*: Assessing the Need for BSSE Corrections

The B3LYP/6-31G* combination performs well and has been widely used for large systems due to its balance of computational efficiency and reasonable accuracy for many molecules. Its popularity is reflected in large-scale projects. Here is how it compares to other common basis sets in terms of computational cost, accuracy, and error compensation

### 1.5.1) Computational Cost

- B3LYP/6-31G* is considered a relatively small basis set, making calculations fast and affordable in terms of CPU time and memory requirements.
- Larger basis sets, such as triple-zeta (e.g., TZVP, 6-311G**, cc-pVTZ) or those with additional polarization and diffuse functions, require substantially more computational resources, often several times more than 6-31G*.
- This efficiency made B3LYP/6-31G* a standard choice for geometry optimizations and property calculations on medium to large molecules, especially when computational resources are limited.

### 1.5.2) Computational Efficiency, Accuracy, and Historical Precedent

- B3LYP/6-31G* provides reasonable accuracy for equilibrium geometries, dipole moments, and vibrational frequencies in medium-sized molecules.
- Qualitative Trends: It reliably predicts qualitative trends and provides reasonable geometries and electronic properties for a broad range of organic molecules, which is useful in screening and exploratory studies.
- Its accuracy for reaction energies and thermochemistry is often the result of error compensation: B3LYP tends to underestimate bond energies, while 6-31G* tends to overbind, so the errors partially cancel out.
- Modern best practice recommends using empirical corrections (such as gCP-D3) or larger, more flexible basis sets (e.g., def2-TZVP, aug-cc-pVTZ) for improved reliability, especially for quantitative work.
- In benchmark studies, split-valence basis sets like 6-31G* often provide similar accuracy to more expensive Dunning basis sets for certain properties, but larger basis sets outperform 6-31G* for challenging cases and when higher accuracy is needed.
- Many databases and comparative studies use this level of theory, facilitating direct comparison and reproducibility across research projects.

#### 1.5.3) Error compensation

It is well established that B3LYP/6-31G* benefits from a fortuitous error compensation between the functional and basis set, which partially offsets the effects of BSSE and missing dispersion. This is discussed in detail by Kruse, Goerigk, and Grimme [45], who recommend the use of empirical corrections (gCP, D3) for improved accuracy, but also note that uncorrected B3LYP/6-31G* often yields reasonable results for organic molecules due to this compensation.

#### 1.5.4) Failure: Missing Dispersion

B3LYP/6-31G* lacks an explicit description of London dispersion (van der Waals) interactions, leading to systematic underestimation of binding energies in noncovalent complexes.

### 1.6) Fundamental challenges

Perfect BSSE correction remains unattainable. CP and CHA have shortcomings. For instance, CP corrections, while widely used, can overcorrect (e.g., decreasing van der Waals overbinding by 10-30% in standard DFT). Modern techniques like rev2 basis sets target BSSE reduction efficiently but at increased computational cost.

## 2) BSSE Corrections in $C_{36}$: Assessing Efficiency and Necessity

B3LYP-gCP/6-31G* computations were carried out on $C_{36}$-$D_{3h}$ and O@$C_{36}$-$D_{3h}$ to determine the influence of BSSE corrections. The data in Tables 8–10 contrast uncorrected and BSSE-corrected results.

### 2.1) Benchmarking B3LYP/6-31G* and B3LYP-gCP-D3/6-31G*: A Quantitative Study and Interpretive Perspective

Table 8 (9) provides the computed bond lengths (valence angles) for the $C_{36}$-$D_{3h}$ and O@$C_{36}$-$D_{3h}$ systems, obtained from B3LYP/6-31G* and B3LYP-gCP/6-31G* calculations.

Table 8: ORCA5 Results for Bond Lengths in $C_{36}(D_{3h})$ and O@$C_{36}(D_{3h})$ Using B3LYP/6-31G and B3LYP-gCP/6-31G. The blue colored values show that O is not centrally positioned in the cage, in that its separating distances from oppositely faced pairs of atoms (14&15 versus 22&25) are not equal.

| $D_{3h}$ symmetry | **Bond length (Å)** | | | |
|---|---|---|---|---|
| | B3LYP/6-31G* | | B3LYP-gCP/ 6-31G* | |
| Definition | $C_{36}$ | O@ $C_{36}$ | $C_{36}$ | O@ $C_{36}$ |
| $C_{14}C_{15}$ | 1.401 | 1.448 | 1.402 | 1.451 |
| $C_{14}C_{17}$ | 1.492 | 1.511 | 1.494 | 1.514 |
| $C_{14}C_{18}$ | 1.445 | 1.496 | 1.447 | 1.499 |
| $C_{15}C_{6}$ | 1.453 | 1.497 | 1.455 | 1.500 |
| $C_{15}C_{11}$ | 1.479 | 1.510 | 1.481 | 1.510 |
| $C_{14}O_{37}$ | - | 1.494 | - | 1.494 |
| $C_{15}O_{37}$ | - | 1.496 | - | 1.495 |
| $C_{25}C_{22}$ | 1.486 | 1.495 | 1.488 | 1.497 |
| $C_{25}C_{29}$ | 1.456 | 1.459 | 1.458 | 1.461 |
| $C_{25}C_{28}$ | 1.395 | 1.382 | 1.396 | 1.384 |
| $C_{22}C_{23}$ | 1.407 | 1.383 | 1.409 | 1.384 |
| $C_{22}C_{19}$ | 1.440 | 1.458 | 1.442 | 1.460 |
| $C_{25}O_{37}$ | - | 3.267 | - | 3.277 |
| $C_{22}O_{37}$ | - | 3.263 | - | 3.275 |

Table 9: ORCA5 Results for Bond Angles in $C_{36}(D_{3h})$ and O@$C_{36}(D_{3h})$ Using B3LYP/6-31G* and B3LYP-gCP/6-31G*

| **Bond angle (°)** | B3LYP/6-31G* | | B3LYP-gCP/6-31G* | |
|---|---|---|---|---|
| Definition | $C_{36}$ | O@$C_{36}$ | $C_{36}$ | O@$C_{36}$ |
| $C_{15}C_{14}C_{17}$ | 119.92 | 122.89 | 119.92 | 122.92 |
| $C_{15}C_{14}C_{18}$ | 118.89 | 120.31 | 118.87 | 120.29 |
| $C_{17}C_{14}C_{18}$ | 107.21 | 113.74 | 107.18 | 113.65 |
| $C_{11}C_{15}C_{14}$ | 120.02 | 123.08 | 120.03 | 123.01 |
| $C_{14}C_{15}C_{6}$ | 119.47 | 120.25 | 119.49 | 120.28 |
| $C_{11}C_{15}C_{6}$ | 106.38 | 113.68 | 106.38 | 113.62 |
| $C_{14}O_{37}C_{15}$ | - | 57.93 | - | 58.05 |
| $C_{25}O_{37}C_{22}$ | - | 26.47 | - | 26.42 |
| $C_{22}C_{25}C_{29}$ | 107.47 | 119.31 | 107.47 | 106.22 |
| $C_{22}C_{25}C_{28}$ | 120.17 | 119.94 | 120.17 | 119.94 |
| $C_{29}C_{25}C_{28}$ | 118.99 | 119.41 | 118.99 | 119.40 |
| $C_{19}C_{22}C_{25}$ | 106.36 | 106.19 | 106.34 | 106.18 |
| $C_{25}C_{22}C_{23}$ | 119.76 | 119.88 | 119.75 | 119.88 |
| $C_{19}C_{22}C_{23}$ | 119.31 | 119.45 | 119.30 | 119.43 |

As shown in Tables 8 and 9, the differences in bond lengths and valence angles between calculations with and without BSSE corrections are negligible. Even for structural features involving the highly electronegative oxygen atom, such as the bond lengths $C_{25}$-$O_{37}$ (0.31%) and $C_{22}$-$O_{37}$ (0.37%), and the angles $C_{14}$-$O_{37}$-$C_{15}$ (0.21%) and $C_{25}$-$O_{37}$-$C_{22}$ (0.19%), the deviations are extremely small.

These findings indicate that applying BSSE corrections has minimal influence on the molecular geometry, emphasizing the structural stability of the molecule irrespective of BSSE adjustments. This indicates that the artificial stabilization arising from basis set overlap does not significantly distort the molecule's potential energy surface or equilibrium geometry. The observed minimal deviations further confirm the computational method's reliability in accurately predicting bond lengths and angles, even in regions containing highly electronegative atoms such as oxygen.

Thus, it can be concluded that BSSE has minimal impact on the geometry of $C_{36}$-$D_{3h}$ and O@$C_{36}$-$D_{3h}$, indicating structural robustness. This stability suggests that both systems behave similarly to rigid structures, a characteristic feature typically observed when non-covalent interactions are weak or negligible.

Table 10 lists key energy parameters, thermodynamic properties, and dipole moments for $C_{36}$-$D_{3h}$ and O@$C_{36}$-$D_{3h}$, calculated with the B3LYP/6-31G* method both with and without BSSE corrections. The most striking observation is that BSSE corrections have no significant effect on the energy-based quantities for the studied systems, which suggests that accurate energetics relations may not require BSSE corrections.

Actually, the data presented demonstrate that the relative differences observed when applying BSSE corrections to the fullerene $C_{36}$-$D_{3h}$ and its endohedral derivative O@$C_{36}$-$D_{3h}$ are negligible for most properties. The total energy shows minimal deviations ($8.04\times10^{-6}$ and $7.63\times10^{-6}$, for the considered fullerene and its derivative respectively), while HOMO/LUMO energies, energy gap, dipole moment, and internal energy exhibit no measurable differences (0.00%). Enthalpy for O@$C_{36}$-$D_{3h}$ displays a moderate deviation (4.87%), whereas entropy and Gibbs free energy show only minor variations (≤0.006% and ≤0.004%, respectively).

These findings indicate that BSSE corrections have a negligible impact on electronic structure properties such as frontier orbital energies, thermodynamic parameters, and the electron density distribution responsible for the dipole moment, suggesting that the chosen basis set adequately minimizes artificial stabilization or distortion effects for these systems. Consequently, BSSE corrections are unnecessary for the reliable analysis of these fullerenes under the current computational framework.

Table 10: ORCA5 Calculations of Energy, Thermodynamics, and Dipole Moment for $C_{36}$-$D_{3h}$ and O@$C_{36}$-$D_{3h}$ with B3LYP/6-31G* and B3LYP-gCP/6-31G*

| $D_{3h}$ symmetry | B3LYP/6-31G* | | B3LYP-gCP/6-31G* | |
|---|---|---|---|---|
| | $C_{36}$ | O@$C_{36}$ | $C_{36}$ | O@$C_{36}$ |
| Total Energy (eV) | -37289.62 | -39330.89 | -37289.61 | -39330.88 |
| $E_{HOMO}$ (eV) | -5.41 | -5.36 | -5.41 | -5.36 |
| $E_{LUMO}$ (eV) | -4.02 | -3.80 | -4.02 | -3.81 |
| $E_{gap}$ (eV) | 1.39 | 1.56 | 1.39 | 1.55 |
| $E_F$ (eV) | -4.72 | -4.58 | -4.72 | -4.59 |
| p (Debye) | 0.079 | 1.011 | 0.079 | 0.68 |
| Inner energy (kcal/mol) | -859772.99 | -906843.62 | -859728.90 | 906792.53 |
| Enthalpy (eV) | -37282.45 | -39323.58 | -37281.61 | -39322.50 |
| Entropy (cal/mol.K) | 109.44 | 110.70 | 107.563 | 110.917 |
| Gibbs energy (eV) | -37283.87 | -39325.02 | -37,283.01 | -39323.50 |

In summary, the observed agreement between geometric parameters and energy values for both $C_{36}$-$D_{3h}$ and O@$C_{36}$-$D_{3h}$ confirms the high reliability of the computational results. This consistency in structural and energetic data underscores the robustness of the findings, demonstrating that the methodology accurately predicts properties for these systems.

Moreover, variations in atomic charges can alter electron density distribution, thereby influencing the magnitude and nature of BSSE, and indeed the BSSE corrections may affect electron density-dependent properties, such as atomic charges and charge distributions. These impacts are particularly significant in charged systems, where careful consideration of BSSE effects is essential for accurate analysis.

Table 11 analyses how BSSE corrections influence atomic charges in the investigated systems, presenting the Mulliken atomic charges for $C_{36}$-$D_{3h}$ and its endohedral derivative O@$C_{36}$-$D_{3h}$, determined using B3LYP/6-31G* and the BSSE-adjusted method.

The data show nearly identical atomic charges for carbons and oxygen, in $C_{36}$-$D_{3h}$ and O@$C_{36}$-$D_{3h}$, obtained with and without BSSE corrections, which highlight BSSE's insignificant impact on electron density distribution.

This suggests that artificial stabilization caused by basis set overlap does not meaningfully distort the computed electron density or the charge distribution among atoms in these systems. As a result, one can conclude that for this inflexible system and basis set, the calculated atomic charges are robust and reliable concerning BSSE corrections, and the choice to apply or omit the CP correction will not meaningfully affect charge-dependent properties or interpretations. The basis set's insufficient flexibility and size indicate that BSSE effects have little impact on the electron density distribution in these systems.

We conclude that the chosen basis set 6-31G*, combined with the rigid structure character of the fullerene cage, is balanced to minimize BSSE effects on electron density-related properties**.**

Remarkably, the oxygen atom's charge density changes minimally, from -0.4818 to -0.4804 (0.29% relative difference), upon BSSE correction. We spotlight in blue color font in both tables 8 and 11 the distances separating O from two oppositely faced pairs of C atoms (pair (14&15) versus (22&25)), and the corresponding charges, interpreting the non-central position of O which approaches the pair (14&15) more than the pair (22&25) due to stronger electrostatic attraction. Figure 6 illustrates the relative position of the O atom, showing furthermore that the pair (14&15) is positioned within hexagons leading to redistribution of stress and thus are more stable than pentagons.

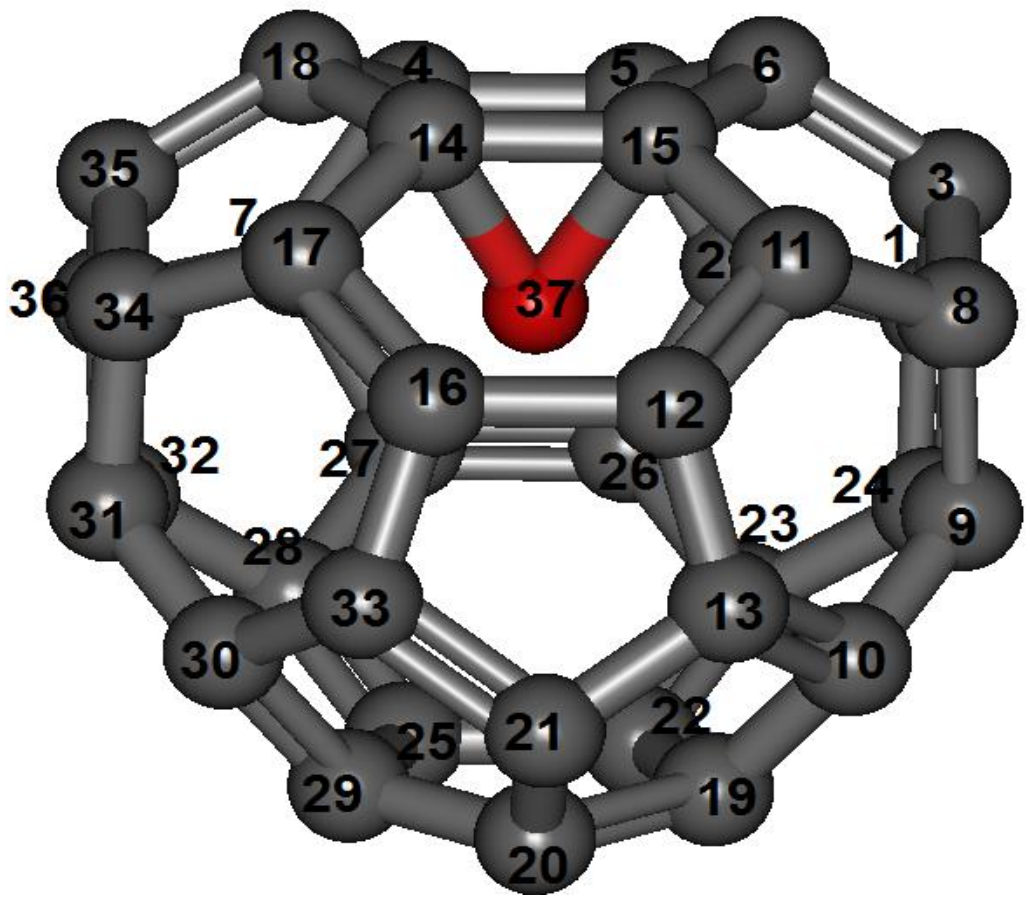

Fig 6: The pair (14,15) faces opposite to the pair (22,25), where O approaches more the former pair due to stronger electrostatic attraction, and due to it being positioned within hexagons increasing stability.

Table 11: Mulliken Atomic Charges for $C_{36}$-$D_{3h}$ and O@$C_{36}$-$D_{3h}$ computed with B3LYP/6-31G* and B3LYP-gCP/6-31G* (ORCA5). The blue colored values conform with O being nearer to the pair (14&15) than the pair (22&25).

| MULLIKEN ATOMIC CHARGE | B3LYP/6-31G* | | B3LYP-gCP/6-31G* | |
|---|---|---|---|---|
| | $C_{36}$ | O@$C_{36}$ | $C_{36}$ | O@$C_{36}$ |
| $C_1$ | -0.0384 | -0.0349 | -0.0385 | -0.0347 |
| $C_2$ | -0.0090 | -0.0076 | -0.0088 | -0.0075 |
| $C_3$ | -0.0363 | -0.0134 | -0.0362 | -0.0133 |
| $C_4$ | 0.0194 | 0.0198 | 0.0192 | 0.0197 |
| $C_5$ | 0.0191 | 0.0190 | 0.0190 | 0.0192 |
| $C_6$ | -0.0108 | 0.0335 | -0.0106 | 0.0333 |
| $C_7$ | -0.0104 | -0.0077 | -0.0102 | -0.0076 |
| $C_8$ | 0.0037 | -0.0145 | 0.0039 | -0.0145 |
| $C_9$ | 0.0159 | 0.0259 | 0.0157 | 0.0258 |
| $C_{10}$ | 0.0188 | 0.0227 | 0.0187 | 0.0224 |
| $C_{11}$ | 0.0184 | 0.0418 | 0.0183 | 0.0417 |
| $C_{12}$ | 0.0168 | 0.0210 | 0.0167 | 0.0212 |
| $C_{13}$ | 0.0006 | -0.0051 | 0.0008 | -0.0049 |
| $C_{14}$ | 0.0158 | 0.1233 | 0.0157 | 0.1229 |
| $C_{15}$ | 0.0197 | 0.1228 | 0.0196 | 0.1224 |
| $C_{16}$ | 0.0195 | 0.0222 | 0.0194 | 0.0218 |
| $C_{17}$ | 0.0177 | 0.0411 | 0.0176 | 0.0412 |
| $C_{18}$ | -0.0090 | 0.0327 | -0.0089 | 0.0329 |
| $C_{19}$ | -0.0197 | 0.0056 | -0.0195 | 0.0058 |
| $C_{20}$ | -0.0350 | -0.0326 | -0.0349 | -0.0326 |
| $C_{21}$ | -0.0349 | -0.0561 | -0.0349 | -0.0561 |
| $C_{22}$ | 0.0200 | 0.0214 | 0.0199 | 0.0211 |
| $C_{23}$ | 0.0179 | 0.0216 | 0.0177 | 0.0215 |
| $C_{24}$ | -0.0171 | -0.0019 | -0.0169 | -0.0017 |
| $C_{25}$ | 0.0161 | 0.0209 | 0.0161 | 0.0207 |
| $C_{26}$ | 0.0154 | 0.0267 | 0.0153 | 0.0264 |
| $C_{27}$ | 0.0202 | 0.0263 | 0.0201 | 0.0261 |
| $C_{28}$ | 0.0182 | 0.0219 | 0.0181 | 0.0217 |
| $C_{29}$ | 0.0005 | 0.0062 | 0.0007 | 0.0063 |
| $C_{30}$ | 0.0189 | 0.0227 | 0.0188 | 0.0224 |
| $C_{31}$ | 0.0162 | 0.0259 | 0.0161 | 0.0258 |
| $C_{32}$ | 0.0034 | -0.0018 | 0.0036 | -0.0016 |
| $C_{33}$ | -0.0199 | -0.0057 | -0.0197 | -0.0054 |
| $C_{34}$ | -0.0170 | -0.0144 | -0.0168 | -0.0145 |
| $C_{35}$ | -0.0385 | -0.0133 | -0.0386 | -0.0133 |
| $C_{36}$ | -0.0364 | -0.0340 | -0.0363 | -0.0341 |
| $O_{37}$ | | -0.4818 | | -0.4804 |
| Average of C \|charge\| | 0.02128 | 0.03113 | 0.018383 | 0.030065 |

### 2.2) Main Issues/implications from the Comparative Analysis

The combination of computational reliability and affordability of B3LYP/6-31G* ensures precise geometry optimization for $C_{36}$-$D_{3h}$ and O@$C_{36}$-$D_{3h}$, even without BSSE corrections. The quasi-rigid nature of these systems reduces susceptibility to BSSE distortions, with weak non-covalent interactions reflected in stable geometries and energies. Electronic properties such as frontier orbitals and dipole moments remain unaffected, and the potential energy surface near equilibrium geometries is unaltered by basis set limitations.

All this confirms B3LYP/6-31G* as a balanced choice for minimizing BSSE effects on structural and electronic properties. The findings validate its application to large and giant fullerenes, including endohedral derivatives, where extended basis sets are computationally expensive enough to be prohibited. Furthermore, non-covalent interactions are expected to diminish in larger fullerenes owing to greater separations between non-bonded atoms, improving the trustworthiness of the B3LYP/6-31G* approach.

## 3) Convergence tests in the ORCA5 program

The convergence tests in ORCA5 are managed using a combination of SCF convergence criteria and geometry optimization parameters to ensure accurate electronic structure and structural outcomes. For rigid systems, stricter convergence tolerances are not essential.

In our work, Geometry optimization was performed using the BFGS quasi-Newton method within redundant internal coordinate frameworks. ORCA5 employs the specified rigorous convergence criteria for the $C_{36}$-$D_{3h}$ and O@$C_{36}$-$D_{3h}$ systems, ensuring reliable predictions of structural, electronic, and energetic properties for these rigid molecular systems.

The convergence settings in ORCA5 for the studied systems are structured into two categories:

1. SCF criteria: Energy tolerance ($10^{-6}$ to $10^{-8}$), density matrix RMS/maximum changes, and orbital-specific thresholds (DIIS error, gradient, rotation).
2. Geometry criteria: Energy/gradient tolerances ($5\times10^{-6}$, $1\times10^{-4}$ Ha/Bohr, $3\times10^{-4}$ Ha/Bohr), displacement limits ($2\times10^{-3}$/$4\times10^{-3}$ Bohr), and Hessian initialization via the Almlöf model.

These settings ensure accuracy for rigid systems like $C_{36}$-$D_{3h}$ and O@$C_{36}$-$D_{3h}$, balancing computational efficiency and precision. The corresponding Convergence tests results, computed using the B3LYP/6-31G* method (uncorrected) and B3LYP-gCP/6-31G* (BSSE-corrected), are detailed in Tables 12–15.

Table 12: ORCA5 Geometry Convergence Tests for $C_{36}$-$D_{3h}$ with B3LYP/6-31G*

| $C_{36}(D_{3h})$: B3LYP/6-31G* Geometry convergence | | | |
|---|---|---|---|
| Item | value | Tolerance | Converged |
| Energy change | -0.0000150686 | 0.0000050000 | NO |
| RMS gradient | 0.0000488493 | 0.0001000000 | YES |
| MAX gradient | 0.0002341599 | 0.0003000000 | YES |
| RMS step | 0.0002727524 | 0.0020000000 | YES |
| MAX step | 0.0008521876 | 0.0040000000 | YES |
| Max(Bonds) 0.0003 Max(Angles) 0.03 | | | |
| Max(Dihed) 0.05 Max(Improp) 0.00 | | | |

For $C_{36}$-$D_{3h}$, the uncorrected method satisfies both RMS and MAX gradient/step criteria, whereas the BSSE-corrected method fails the MAX gradient test. Despite this, tolerance thresholds remain consistent regardless of BSSE inclusion. Identical convergence behavior is observed for O@$C_{36}$-$D_{3h}$, where BSSE corrections similarly disrupt the MAX gradient criterion without altering tolerance limits.

Table 13: ORCA5 Geometry Convergence Tests for $C_{36}$-$D_{3h}$ with B3LYP-gCP/6-31G*

| $C_{36}$($D_{3h}$): B3LYP-gCP /6-31G* Geometry convergence | | | |
|---|---|---|---|
| Item | value | Tolerance | Converged |
| Energy change | -0.0000873617 | 0.0000050000 | NO |
| RMS gradient | 0.0000723971 | 0.0001000000 | YES |
| MAX gradient | 0.0003084300 | 0.0003000000 | NO |
| RMS step | 0.0002531171 | 0.0020000000 | YES |
| MAX step | 0.0008952662 | 0.0040000000 | YES |
| Max(Bonds) 0.0005 Max(Angles) 0.02 | | | |
| Max(Dihed) 0.04 Max(Improp) 0.00 | | | |

Table 14: ORCA5 Geometry Convergence Tests for O@$C_{36}$-$D_{3h}$ with B3LYP/6-31G*

| O@$C_{36}$($D_{3h}$): B3LYP/6-31G* Geometry convergence | | | |
|---|---|---|---|
| Item | value | Tolerance | Converged |
| Energy change | -0.0000378293 | 0.0000050000 | NO |
| RMS gradient | 0.0000333355 | 0.0001000000 | YES |
| MAX gradient | 0.0002642297 | 0.0003000000 | YES |
| RMS step | 0.0005059377 | 0.0020000000 | YES |
| MAX step | 0.0017363377 | 0.0040000000 | YES |
| Max(Bonds) 0.0008 Max(Angles) 0.07 | | | |
| Max(Dihed) 0.10 Max(Improp) 0.00 | | | |

Table 15: ORCA5 Geometry Convergence Tests for O@$C_{36}$-$D_{3h}$ with B3LYP-gCP/6-31G*

| O@$C_{36}$($D_{3h}$): B3LYP-gCP /6-31G* Geometry convergence | | | |
|---|---|---|---|
| Item | value | Tolerance | Converged |
| Energy change | -0.0000071921 | 0.0000050000 | NO |
| RMS gradient | 0.0000367321 | 0.0001000000 | YES |
| MAX gradient | 0.0002009789 | 0.0003000000 | YES |
| RMS step | 0.0002565182 | 0.0020000000 | YES |
| MAX step | 0.0010458887 | 0.0040000000 | YES |
| Max(Bonds) 0.0005 Max(Angles) 0.03 | | | |
| Max(Dihed) 0.06 Max(Improp) 0.00 | | | |

**References**

[1] Thilgen, C. and Diederich, F., 2006. Structural aspects of fullerene chemistry a journey through fullerene chirality. Chemical reviews, 106(12), pp.5049-5135.
[2] Manna, D. and Ghanty, T. K. 2013. Enhancement in the Stability of 36-Atom Fullerene through Encapsulation of a Uranium Atom, Journal of Physical Chemistry. C 2013, 117(34), pp. 17859−17869.
[3] Popov, A.A., Yang, S. and Dunsch, L., 2013. Endohedral fullerenes. Chemical reviews, 113(8), pp.5989-6113.
[4] Dunk, P.W., Kaiser, N.K., Mulet-Gas, M., Rodríguez-Fortea, A., Poblet, J.M., Shinohara, H., Hendrickson, C.L., Marshall, A.G. and Kroto, H.W., 2012. The smallest stable fullerene, M@ C28 (M= Ti, Zr, U):

stabilization and growth from carbon vapor. Journal of the American Chemical Society, 134(22), pp.9380-9389.
[5] Naderi, F., Rostamian, S. and Naderi, B., 2012. A study on the electronic and structural properties of fullerene C36 and it's interaction with amino acid. International Journal of Physical Sciences, 7(13), pp.2006-2009.
[6] Piskoti, C., Yarger, J. and Zettl, A., 1998. C36, a new carbon solid. nature, 393(6687), pp.771-774.
[7] Fowler, P.W. and Manolopoulos, D.E., 2007. An atlas of fullerenes. Courier Corporation.
[8] Piskoti, C. and Zettl, A., 1998, August. The first stable lower fullerene: C 36. In AIP Conference Proceedings (Vol. 442, No. 1, pp. 183-185). American Institute of Physics.
[9] Grossman, J.C., Côté, M., Louie, S.G. and Cohen, M.L., 1998. Electronic and structural properties of molecular C36. Chemical physics letters, 284(5-6), pp.344-349.
[10] Côté, M., Grossman, J.C., Cohen, M.L. and Louie, S.G., 1998. Electron-phonon interactions in solid C 36. Physical Review Letters, 81(3), p.697.
[11] Halac, E., Burgos, E. and Bonadeo, H., 1999. Molecular structure and dynamical properties of C36: a semi-empirical calculation. Chemical physics letters, 299(1), pp.64-68.
[12] Fowler, P.W., Heine, T., Rogers, K.M., Sandall, J.P.B., Seifert, G. and Zerbetto, F., 1999. C36, a hexavalent building block for fullerene compounds and solids. Chemical physics letters, 300(3-4), pp.369-378.
[13] Jishi, R.A. and Dresselhaus, M.S., 1999. Vibrational frequencies in C36. Chemical physics letters, 302(5-6), pp.533-537.
[14] Ito, A., Monobe, T., Yoshii, T. and Tanaka, K., 1999. Electronic structures of C36 fulleride anions: C36− and C362−. Chemical Physics Letters, 315(5-6), pp.348-354.
[15] Beu, T.A., Onoe, J. and Takeuchi, K., 2001. Structural and vibrational properties of C 36 and its oligomers (C 36) M= 2, 3, 4 by tight-binding molecular dynamics. The European Physical Journal D-Atomic, Molecular, Optical and Plasma Physics, 17, pp.205-212.
[16] Varganov, S.A., Avramov, P.V., Ovchinnikov, S.G. and Gordon, M.S., 2002. A study of the isomers of C36 fullerene using single and multireference MP2 perturbation theory. Chemical physics letters, 362(5-6), pp.380-386.
[17] Paulus, B., 2003. Electronic and structural properties of the cage-like molecules C 20 to C 36. Physical Chemistry Chemical Physics, 5(16), pp.3364-3367.
[18] Jin, Y.F. and Hao, C., 2005. Computational Study of the Stone− Wales Transformation in C36. The Journal of Physical Chemistry A, 109(12), pp.2875-2877.
[19] Kim, K.H., Han, Y.K. and Jung, J., 2005. Basis set effects on relative energies and HOMO–LUMO energy gaps of fullerene C36. Theoretical Chemistry Accounts, 113(4), pp.233-237.
[20] Małolepsza, E., Witek, H.A. and Irle, S., 2007. Comparison of geometric, electronic, and vibrational properties for isomers of small fullerenes C20− C36. The Journal of Physical Chemistry A, 111(29), pp.6649-6657.
[21] Ma, Y., Zhang, J.R., Wang, S.Y., Hu, J., Lin, J. and Song, X.N., 2019. Distinguishing the six stable C36 fullerene isomers by means of soft X-ray spectroscopies at DFT level. Molecular Physics, 117(5), pp.635-643.
[22] Idrissi, S., Jabar, A. and Bahmad, L., 2024. Study of magnetic properties of the fullerene C36 structure by Monte Carlo simulations. Indian Journal of Physics, pp.1-8.
[23] Dunk, P.W., Mulet-Gas, M., Nakanishi, Y., Kaiser, N.K., Rodríguez-Fortea, A., Shinohara, H., Poblet, J.M., Marshall, A.G. and Kroto, H.W., 2014. Bottom-up formation of endohedral mono-metallofullerenes is directed by charge transfer. Nature Communications, 5(1), p.5844.
[24] Klingeler, R., Bechthold, P.S., Neeb, M. and Eberhardt, W., 2000. Mass spectra of metal-doped carbon and fullerene clusters. The Journal of Chemical Physics, 113(4), pp.1420-1425.
[25] LI, J., Huang, X., Wu, L., Zhang, Y. and Wu, K., 2000. DFT studies on the electronic structure of C36 (D6h symmetry) and its derivatives. Acta Chimica Sinica, 58(3), p.319.

[26] Huang, X., Li, J.Q. and Zhang, Y.F., 2000. DFT studies on the electronic structure of C-36 (D-6h symmetry) and its alkaline earths metals derivatives. CHINESE JOURNAL OF STRUCTURAL CHEMISTRY, 19(1), pp.64-68.
[27] Slanina, Z. and Chow, T.J., 2003. He and N encapsulations in small cages: Model computations for possible nanotechnology systems. Journal of Nanoscience and Nanotechnology, 3(4), pp.303-307.
[28] Wang, L., Han, P., Jia, H. and Xu, B. 2006. Influence of Mo atom dope on structure and properties of C36, Cailiao Yanjiu Xuebao/Chinese Journal of Materials Research 20(4), pp.431-434.
[29] Kang, H.S., 2006. Theoretical study of endohedral C36 and its dimers. The Journal of Physical Chemistry A, 110(14), pp.4780-4786.
[30] Kim, Y.H., Zhao, Y., Williamson, A., Heben, M.J. and Zhang, S.B., 2006. Nondissociative adsorption of H 2 molecules in light-element-doped fullerenes. Physical Review Letters, 96(1), p.016102.
[31] Garg, I., Sharma, H., Kapila, N., Dharamvir, K. and Jindal, V.K., 2011. Transition metal induced magnetism in smaller fullerenes (C n for n≤ 36). Nanoscale, 3(1), pp.217-224.
[32] Sure, R., Tonner, R. and Schwerdtfeger, P., 2015. A systematic study of rare gas atoms encapsulated in small fullerenes using dispersion corrected density functional theory. Journal of Computational Chemistry, 36(2), pp.88-96.
[33] Lee, T.G., Ludlow, J.A. and Pindzola, M.S., 2012. Single photoionization with excitation and double photoionization of He@ C36, He@ C60 and He@ C82. Journal of Physics B: Atomic, Molecular and Optical Physics, 45(13), p.135202.
[34] Miralrio, A. and Enrique Sansores, L., 2017. Structures, stabilities, and electronic properties of fullerene C36 with endohedral atomic Sc, Y, and La: A dispersion-corrected DFT study. International Journal of Quantum Chemistry, 117(6), p.e25335.
[35] Harneit, W., Meyer, C., Weidinger, A., Suter, D. and Twamley, J., 2002. Architectures for a spin quantum computer based on endohedral fullerenes. physica status solidi (b), 233(3), pp.453-461.
[36] Madhu Menon and Ernst Richter, Structure and stability of Solid $C_{36}$ . Phys. Rev. B **60**, 13322 (1999)
[37] Kerim, A. Aromaticity and kinetic stability of fullerene $C_{36}$ isomers and their molecular ions. *J Mol Model* **17**, 3257–3263 (2011). https://doi.org/10.1007/s00894-011-1012-9
[38] Neese, F., 2022. Software update: The ORCA program system—Version 5.0. Wiley Interdisciplinary Reviews: Computational Molecular Science, 12(5), p.e1606.
[39] Hobza, Pavel; Müller-Dethlefs, Klaus (2010). *Non-covalent Interactions: Theory and Experiment*. Cambridge, England: Royal Society of Chemistry.
[40] Frans B. van Duijneveldt, Jeanne G. C. M. van Duijneveldt-van de Rijdt, and Joop H. van Lenthe. State of the Art in Counterpoise Theory**.** *Chemical Reviews* **1994** *94* (7), 1873-1885**,** DOI: 10.1021/cr00031a007
[41] ORCA 6.0 manual, ORCA Input Library: https://sites.google.com/site/orcainputlibrary/geometry-input

[42] Kuchitsu, K. ed., 1995. Landolt-Börnstein: Molecules and Radicals. Structure Data of Free Polyatomic Molecules: Supplement to Volume II/7, II/15, II/21/Ed.: K. Kuchitsu. Contributors: G. Graner... Springer.
[43] Azam, F., Alabdullah, N.H., Ehmedat, H.M., Abulifa, A.R., Taban, I. and Upadhyayula, S., 2018. NSAIDs as potential treatment option for preventing amyloid β toxicity in Alzheimer's disease: an investigation by docking, molecular dynamics, and DFT studies. Journal of Biomolecular Structure and Dynamics, 36(8), pp.2099-2117.
[44] Schwerdtfeger, P. and Nagle, K.J., 2019. Table of static dipole polarizabilities of the neutral elements in the periodic table (vol 117, pg 1200, 2018). Mol. Phys, 117, p.1585.
[45] H. Kruse, L. Goerigk, and S. Grimme 2012, Why the Standard B3LYP/6-31G* Model Chemistry Should Not Be Used in DFT Calculations of Molecular Thermochemistry: Understanding and Correcting the Problem. Journal of Organic Chemistry, 77, 10824-10834